\documentclass[pra,aps,superscriptaddress,amssymb,amsmath,reprint,noeprint,floatfix]{revtex4-1}
\usepackage{natbib}
\usepackage{braket}
\usepackage{bm}
\usepackage{graphicx}
\usepackage{hyperref}
\usepackage{xcolor}
\hypersetup{bookmarksnumbered}
\newcommand{\uvec}[1]{\boldsymbol{\hat{\mathbf{#1}}}}
\newcommand{\Tvec}[1]{\tensor{\mathbf{#1}}}

\newcommand{\EqnOne}{
    \begin{equation}
        \braket{\mathbf{F}}=\oint_S\braket{\Tvec{T}}\cdot d\mathbf{A},
        \label{eqn:1}
    \end{equation}
}

\newcommand{\EqnTwo}{
    \begin{equation}
        \begin{aligned}
        \braket{T_{ij}}=\frac{1}{2}\mathrm{Re}&\left[ \varepsilon_0E_iE_j^*+\frac{1}{\mu_0}B_iB_j^*\right.\\
        &-\left.\frac{1}{2}\left( \varepsilon_0|\mathbf{E}|^2+\frac{1}{\mu_0}|\mathbf{B}|^2\right)\delta_{ij}\right]
        \end{aligned}
        \label{eqn:2}
    \end{equation}
}

\newcommand{\EqnThree}{
    \begin{equation}
        p_i=-\frac{1}{i\omega}\int_VJ_i\,dv',\quad m_i=\frac{1}{2}\int_V\left( \mathbf{r}'\times\mathbf{J}\right)_i\,dv',
        \label{eqn:3}
    \end{equation}
}

\newcommand{\EqnFour}{
    \begin{equation}
        \begin{aligned}
            \mathbf{F^p}&=\frac{1}{2}\mathrm{Re}\left[ \left( \nabla \mathbf{E}^*\right)\cdot \mathbf{p}\right],\\
            \mathbf{F^m}&=\frac{1}{2}\mathrm{Re}\left[ \left( \nabla \mathbf{B}^*\right)\cdot \mathbf{m}\right],\\
            \mathbf{F}^\mathrm{int}&=-\frac{k^4}{12\pi\varepsilon_0c}\mathrm{Re}\left[ \mathbf{p}\times\mathbf{m}^*\right].
        \end{aligned}
        \label{eqn:4}
    \end{equation}
}

\newcommand{\EqnFive}{
    \begin{equation}
        \begin{aligned}
            \mathbf{E}_\mathrm{sca}^\mathbf{p}&=\frac{1}{4\pi\varepsilon_0}\left\{ k^2\left( \mathbf{n}\times\mathbf{p}\right)\times\mathbf{n}\frac{e^{ikr}}{r}\right.
            \\&\quad
            \left.+\left[ 3\mathbf{n}\left( \mathbf{n}\cdot\mathbf{p}\right)-\mathbf{p}\right]\left( \frac{1}{r^3}-\frac{ik}{r^2}\right)e^{ikr}\right\},\protect\\
            \mathbf{B}_\mathrm{sca}^\mathbf{p}&=\frac{\mu_0ck^2}{4\pi}\left( \mathbf{n}\times\mathbf{p}\right)\frac{e^{ikr}}{r}\left( 1-\frac{1}{ikr}\right),\protect\\
            \mathbf{E}_\mathrm{sca}^\mathbf{m}&=-\frac{Z_0k^2}{4\pi}\left( \mathbf{n}\times\mathbf{m}\right)\frac{e^{ikr}}{r}\left( 1-\frac{1}{ikr}\right),\protect\\
            \mathbf{B}_\mathrm{sca}^\mathbf{m}&=\frac{\mu_0}{4\pi}\left\{ k^2\left( \mathbf{n}\times\mathbf{m}\right)\times\mathbf{n}\frac{e^{ikr}}{r}\right.
            \protect\\&\quad
            \left.+\left[ 3\mathbf{n}\left( \mathbf{n}\cdot\mathbf{m}\right)-\mathbf{m}\right]\left( \frac{1}{r^3}-\frac{ik}{r^2}\right)e^{ikr}\right\},\protect\\
        \end{aligned}
        \label{eqn:5}
    \end{equation}
}

\newcommand{\EqnSix}{
    \begin{equation}
        \begin{aligned}
            F_i^\mathbf{p}&=\frac{1}{2}\mathrm{Re}\left[ p_j\partial_iE_j^*\right],\protect\\
            F_i^\mathbf{m}&=\frac{1}{2}\mathrm{Re}\left[ m_j\partial_iB_j^*\right],\protect\\
            F_i^\mathrm{int}&=-\frac{k^4}{12\pi\varepsilon_0c}\epsilon_{ijk}\mathrm{Re}\left[ p_jm_k^*\right].
        \end{aligned}
        \label{eqn:6}
    \end{equation}
}

\newcommand{\EqnSeven}{
    \begin{equation}
        \begin{aligned}
        \sigma_\mathrm{sca}^\mathbf{p}&=\frac{1}{S_\mathrm{inc}}\oint_S \frac{1}{2\mu_0} \mathrm{Re} \left[ \mathbf{E}^\mathbf{p}_\mathrm{sca}\times \mathbf{B}^\mathbf{p\, *}_\mathrm{sca}\right]\cdot d\mathbf{A},\protect\\
        \sigma_\mathrm{sca}^\mathbf{m}&=\frac{1}{S_\mathrm{inc}}\oint_S \frac{1}{2\mu_0} \mathrm{Re} \left[ \mathbf{E}^\mathbf{m}_\mathrm{sca}\times \mathbf{B}^\mathbf{m\, *}_\mathrm{sca}\right]\cdot d\mathbf{A}.
        \end{aligned}
        \label{eqn:7}
    \end{equation}
}

\newcommand{\EqnEightold}{
        \begin{equation}
            \begin{aligned}
                \frac{dA_\mathbf{p}^\mathcal{T}}{dt}=&-i\omega_0A_\mathbf{p}^\mathcal{T}-\Gamma_\mathrm{e}A_\mathbf{p}^\mathcal{T}-i\kappa_\mathrm{ee}A_\mathbf{p}^\mathcal{B}+\sqrt{\gamma_\mathrm{e}}S_\mathrm{TE}^\mathcal{T}\\
                &+\sqrt{\gamma_\mathrm{em}}S_\mathrm{TM}^\mathcal{T},\protect\\
                \frac{dA_\mathbf{p}^\mathcal{B}}{dt}=&-i\omega_0A_\mathbf{p}^\mathcal{B}-\Gamma_\mathrm{e}A_\mathbf{p}^\mathcal{B}-i\kappa_\mathrm{ee}A_\mathbf{p}^\mathcal{T}+\sqrt{\gamma_\mathrm{e}}S_\mathrm{TE}^\mathcal{B}\\
                &+\sqrt{\gamma_\mathrm{em}}S_\mathrm{TM}^\mathcal{B},\protect\\
                \frac{dA_\mathbf{m}^\mathcal{T}}{dt}=&-i\omega_0A_\mathbf{m}^\mathcal{T}-\Gamma_\mathrm{m}A_\mathbf{m}^\mathcal{T}-i\kappa_\mathrm{mm}A_\mathbf{m}^\mathcal{B}+\sqrt{\gamma_\mathrm{em}}S_\mathrm{TE}^\mathcal{T}\\
                &+\sqrt{\gamma_\mathrm{m}}S_\mathrm{TM}^\mathcal{T},\protect\\
                \frac{dA_\mathbf{m}^\mathcal{B}}{dt}=&-i\omega_0A_\mathbf{m}^\mathcal{B}-\Gamma_\mathrm{m}A_\mathbf{m}^\mathcal{B}-i\kappa_\mathrm{mm}A_\mathbf{m}^\mathcal{T}+\sqrt{\gamma_\mathrm{em}}S_\mathrm{TE}^\mathcal{B}\\
                &+\sqrt{\gamma_\mathrm{m}}S_\mathrm{TM}^\mathcal{B},\protect\\
                 \label{eqn:8}
            \end{aligned}
        \end{equation}
}

\newcommand{\EqnNine}{
    \begin{equation}
        \frac{d\Lambda}{dt}=-iH\Lambda+CS
        \label{eqn:9}
    \end{equation}
}

\newcommand{\EqnTen}{
    \begin{equation}
        \begin{aligned}
            H&=
            \begin{bmatrix}
                \omega_0-i\Gamma_\mathrm{e} && \kappa_\mathrm{ee} &&&&\protect\\
                \kappa_\mathrm{ee} && \omega_0-i\Gamma_\mathrm{e} &&&&\protect\\
                
                &&&&\omega_0-i\Gamma_\mathrm{m} && \kappa_\mathrm{mm}\protect\\
                &&&&\kappa_\mathrm{mm} && \omega_0-i\Gamma_\mathrm{m}\protect\\
            \end{bmatrix},\protect\\
            \Lambda&=
            \begin{bmatrix}
                A_\mathbf{p}^\mathcal{T}\protect\\
                A_\mathbf{p}^\mathcal{B}\protect\\
                A_\mathbf{m}^\mathcal{T}\protect\\
                A_\mathbf{m}^\mathcal{B}\protect\\
            \end{bmatrix},
            \quad
            C=
            \begin{bmatrix}
                \sqrt{\gamma_\mathrm{e}}&&\sqrt{\gamma_\mathrm{em}}\protect\\
                \sqrt{\gamma_\mathrm{e}}&&\sqrt{\gamma_\mathrm{em}}\protect\\
                \sqrt{\gamma_\mathrm{em}}&&\sqrt{\gamma_\mathrm{m}}\protect\\
                \sqrt{\gamma_\mathrm{em}}&&\sqrt{\gamma_\mathrm{m}}\protect\\
            \end{bmatrix},\protect\\
            S&=
            \begin{bmatrix}
                S_\mathrm{TE}^\mathcal{T}&&S_\mathrm{TE}^\mathcal{B}&&S_\mathrm{TE}^\mathcal{T}&&S_\mathrm{TE}^\mathcal{B}\protect\\
                S_\mathrm{TM}^\mathcal{T}&&S_\mathrm{TM}^\mathcal{B}&&S_\mathrm{TM}^\mathcal{T}&&S_\mathrm{TM}^\mathcal{B}
            \end{bmatrix}.
        \end{aligned}
        \label{eqn:10}
    \end{equation}
}

\newcommand{\EqnEleven}{
    \newcommand{\numeratorBp}{
        i\sqrt{\gamma_\mathrm{em}}\left( S_\mathrm{TE}^\mathcal{T}\kappa + S_\mathrm{TE}^\mathcal{B}\left( i\Gamma_\mathrm{e}+ \omega-\omega_0\right)\right)
    }
    \newcommand{\numeratorTp}{
        i\sqrt{\gamma_\mathrm{em}}\left( S_\mathrm{TE}^\mathcal{B}\kappa+S_\mathrm{TE}^\mathcal{T}\left( i\Gamma_\mathrm{e}+\omega-\omega_0\right)\right)
    }
    \newcommand{\numeratorBm}{
        i\sqrt{\gamma_\mathrm{m}}\left( S_\mathrm{TM}^\mathcal{T}\kappa + S_\mathrm{TM}^\mathcal{B}\left( i\Gamma_\mathrm{m}+ \omega-\omega_0\right)\right)
    }
    \newcommand{\numeratorTm}{
        i\sqrt{\gamma_\mathrm{m}}\left( S_\mathrm{TM}^\mathcal{B}\kappa+S_\mathrm{TM}^\mathcal{T}\left( i\Gamma_\mathrm{m}+\omega-\omega_0\right)\right)
    }
    
    \newcommand{\denominatorp}{
        \kappa^2-\left( i\Gamma_\mathrm{e} +  \omega-\omega_0\right)^2
    }
    \newcommand{\denominatorm}{
        \kappa^2-\left( i\Gamma_\mathrm{m} +  \omega-\omega_0\right)^2
    }
    \begin{equation}
        \begin{aligned}
            A_\mathbf{p}^\mathcal{B}=&-\frac{\numeratorBp}{\denominatorp},\protect\\
            A_\mathbf{p}^\mathcal{T}=&-\frac{\numeratorTp}{\denominatorp},\protect\\
            A_\mathbf{m}^\mathcal{B}=&-\frac{\numeratorBm}{\denominatorm},\protect\\
            A_\mathbf{m}^\mathcal{T}=&-\frac{\numeratorTm}{\denominatorm}.
        \end{aligned}
        \label{eqn:11}
    \end{equation}
}

\newcommand{\FigOne}{
    \begin{figure}[t!]
        \centering
        \includegraphics[width=0.7\linewidth]{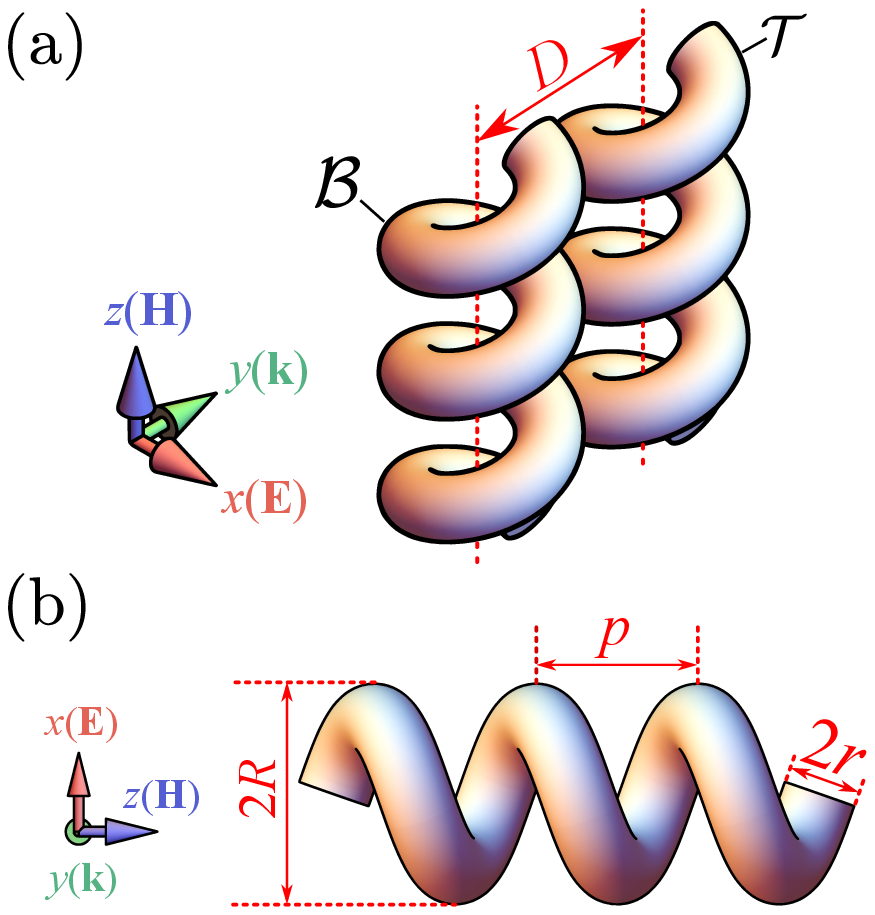}
        \caption{(a) Drawn-to scale schematic of the model system. Two gold helices are under the illumination of an electromagnetic plane wave propagating in $y$ direction and polarized in $x$ direction.(b) Size view of the model system.}
        \label{fig:1}
    \end{figure}
}

\newcommand{\FigTwo}{
    \begin{figure}[t!]
        \centering
        \includegraphics[width=0.8\linewidth]{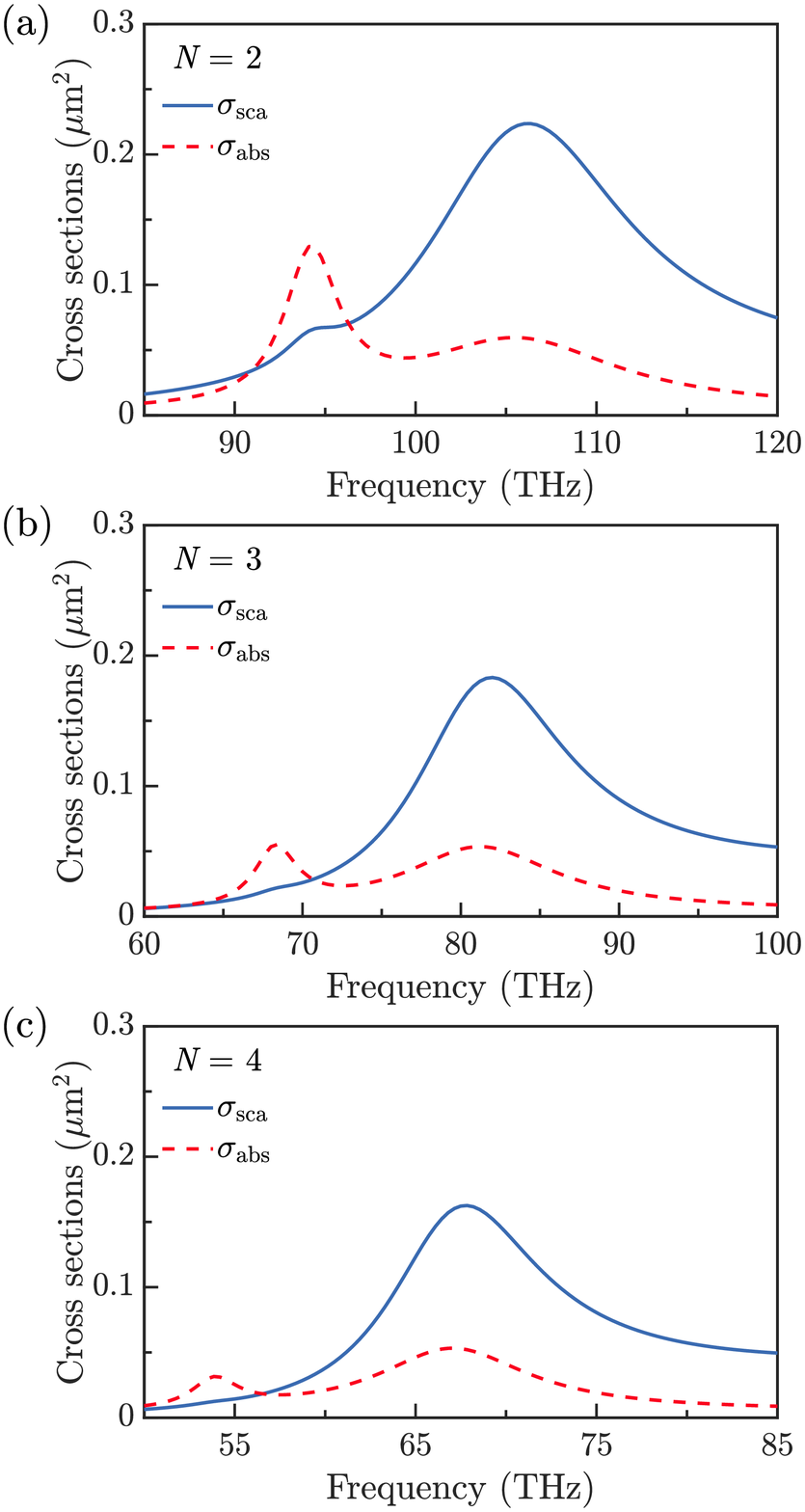}
        \caption{Scattering and absorption cross sections of coupled chiral particles with pitch number (a) $N=2$, (b) $N=3$, (c) $N=4$. The  distance is fixed at $D= 340$ nm.}
        \label{fig:2abc}
    \end{figure}
}

\newcommand{\FigThree}{
    \begin{figure}[t!]
        \centering
        \includegraphics[width=0.8\linewidth]{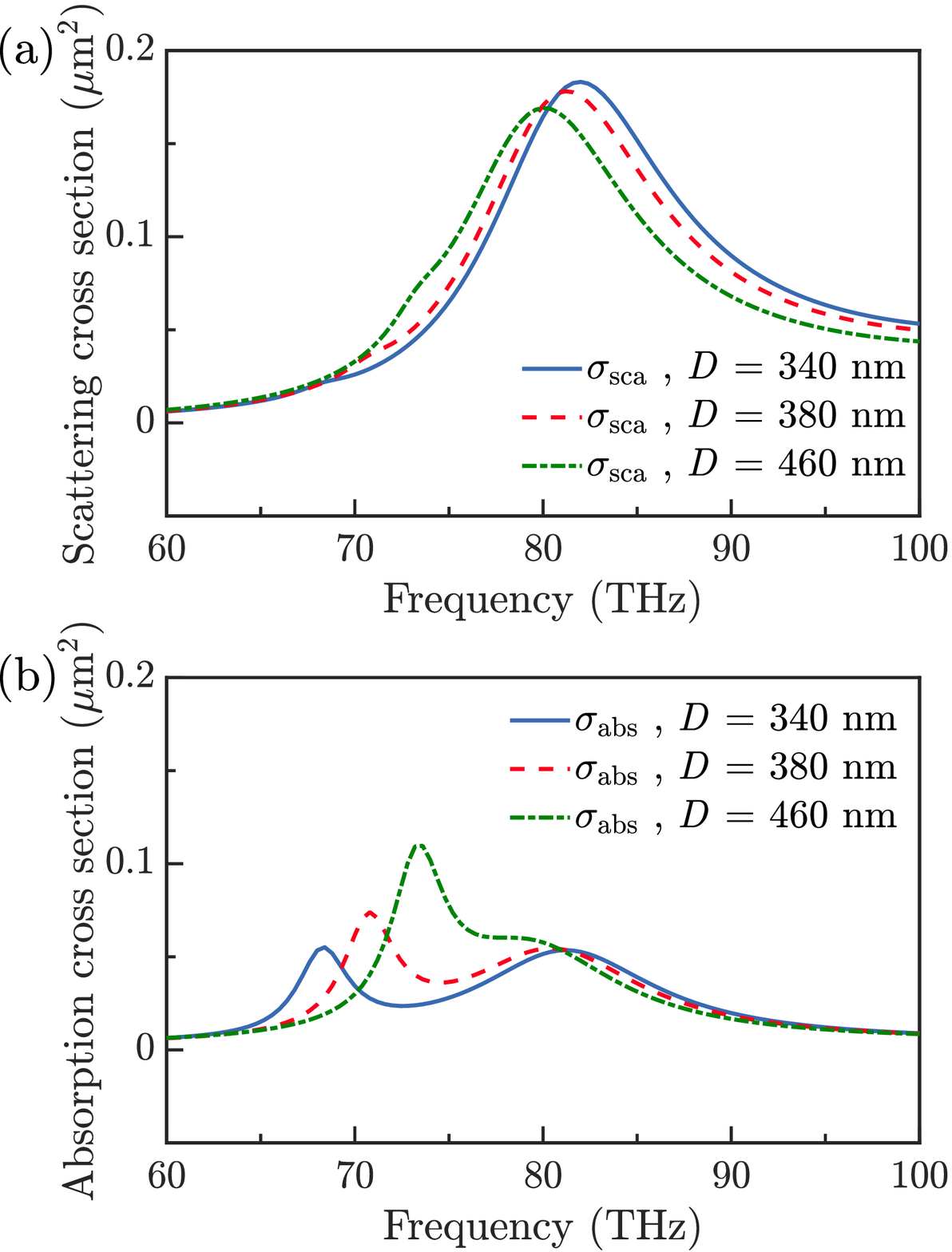}
        \caption{Dependence of the (a) scattering cross section and (b) absorption cross section on the coupling distance $D$. We fix $N=3$.}
        \label{fig:3ab}
    \end{figure}
}

\newcommand{\FigFour}{
    \begin{figure*}[t!]
        \centering
        \includegraphics[width=\linewidth]{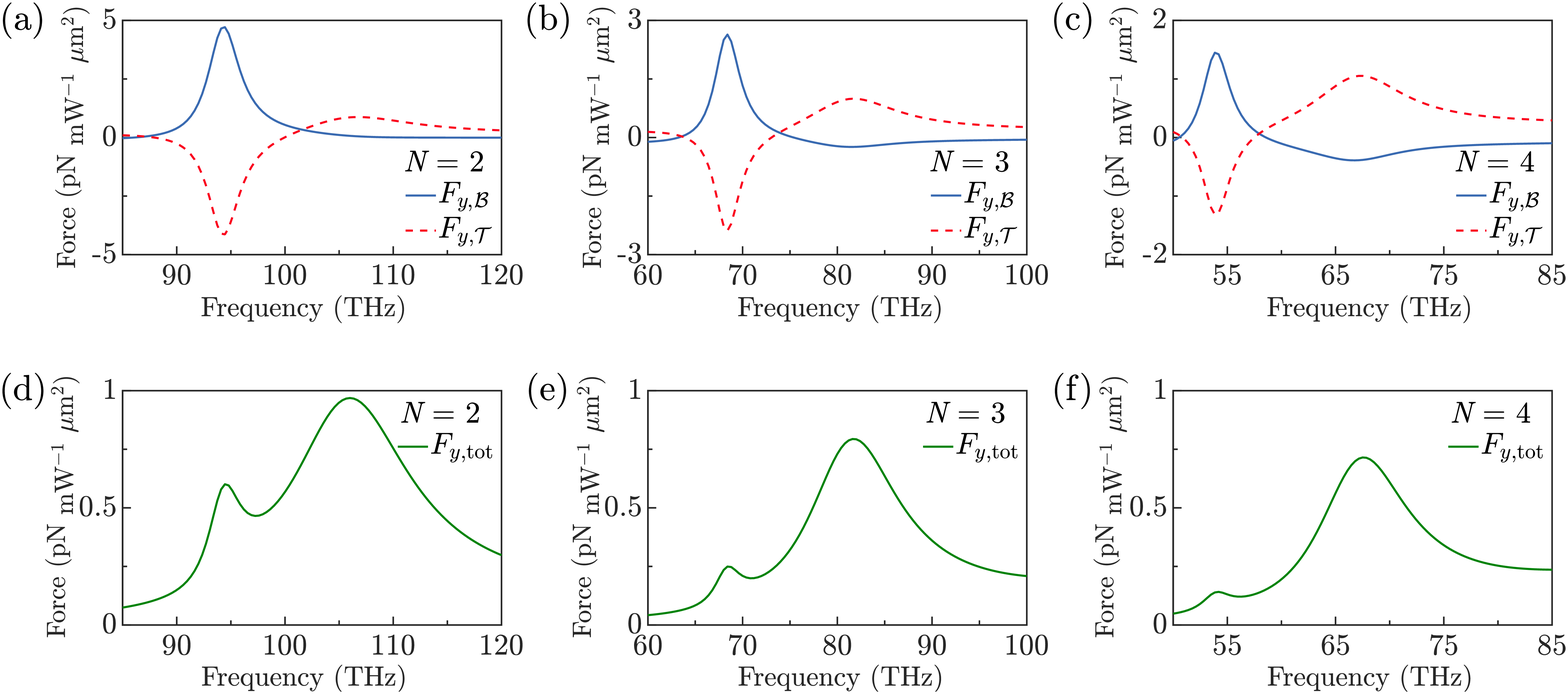}
        \caption{Optical forces on the top helix and the bottom helix [(a), (b), (c)], and the whole system [(d), (e), (f)] for different pitch numbers $N=2,3,4$.}
        \label{fig:4abcdefghi}
    \end{figure*}
}

\newcommand{\FigFive}{
    \begin{figure}[t!]
        \centering
        \includegraphics[width=0.8\linewidth]{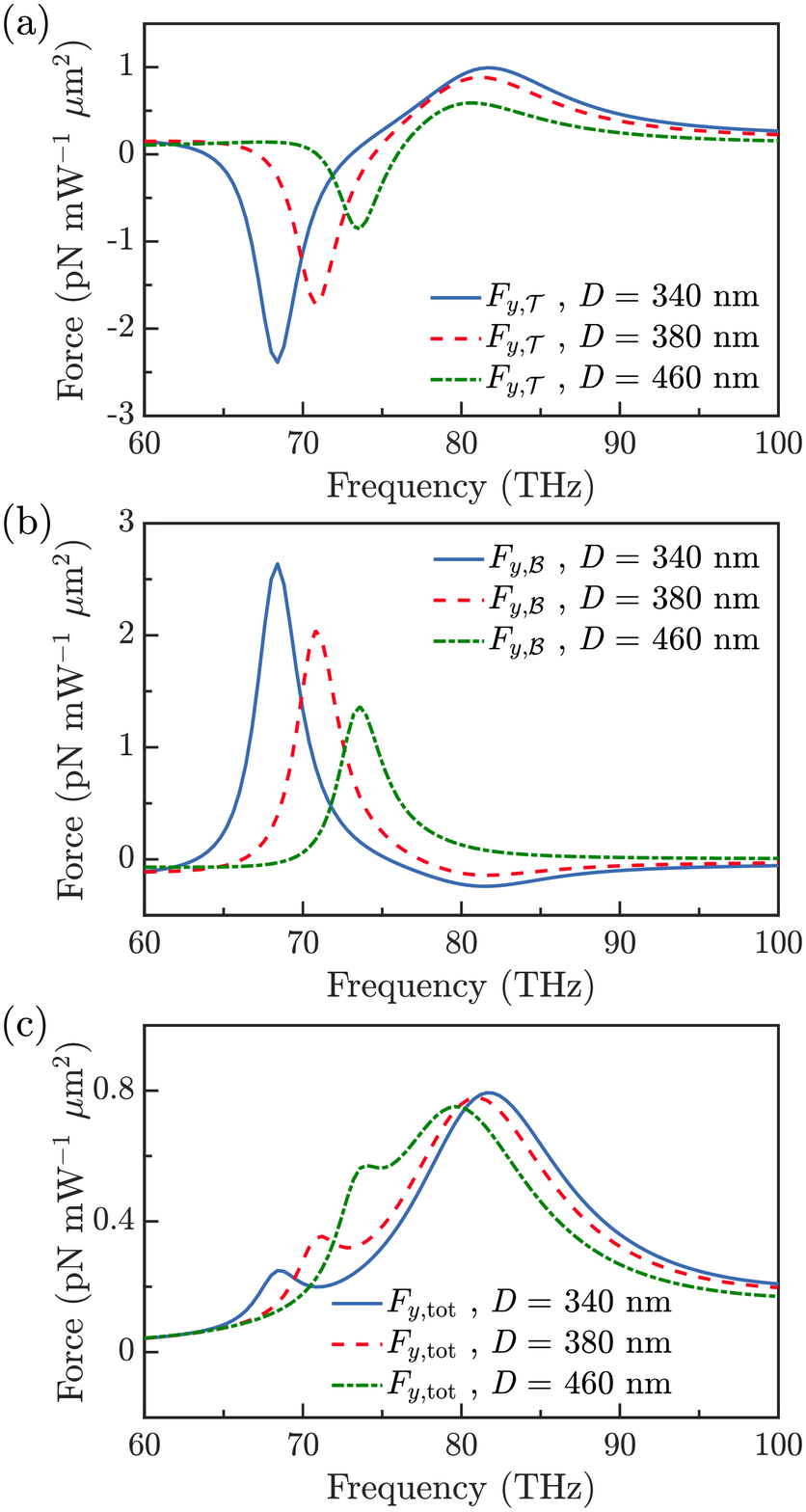}
        \caption{Optical force on the (a) top helix, (b) bottom helix, and (c) whole system for different coupling distance $D$. We fix $N=3$.}
        \label{fig:5abc}
    \end{figure}
}

\newcommand{\FigSix}{
    \begin{figure}[tb!]
        \centering
        \includegraphics[width=0.8\linewidth]{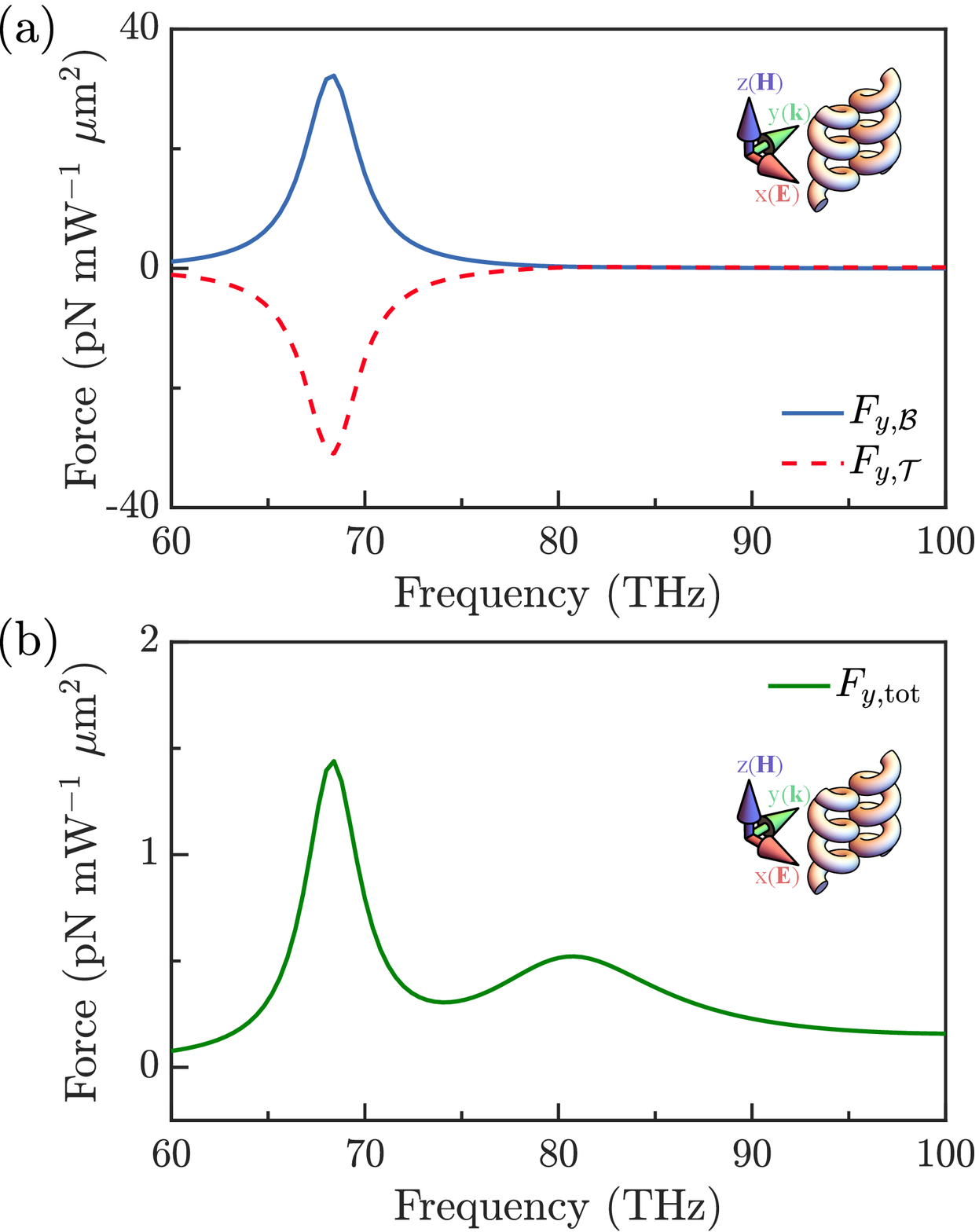}
        \caption{Optical force on the (a) top and bottom helices and (b) whole system in coupled chiral particles with opposite handedness.}
        \label{fig:6abc}
    \end{figure}
}

\newcommand{\FigSeven}{
    \begin{figure}[tb!]
        \centering
        \includegraphics[width=0.8\linewidth]{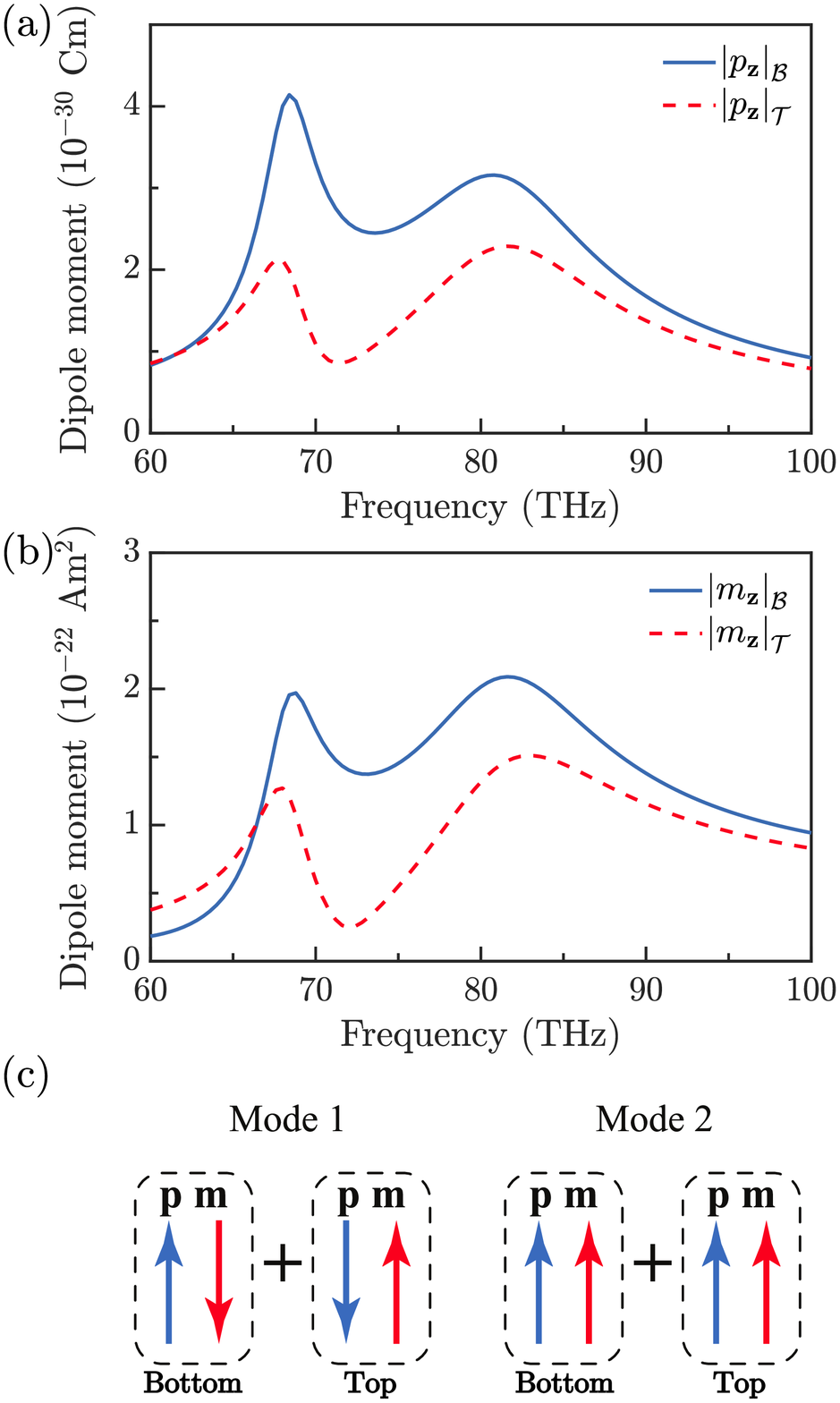}
        \caption{(a) Electric dipole moments and (b) magnetic dipole moments induced in the coupled chiral particles of same handedness. (c) Schematics showing the direction of the induced dipole moments at the two resonances.}
        \label{fig:7abc}
    \end{figure}
}

\newcommand{\FigEight}{
    \begin{figure}[tb!]
        \centering
        \includegraphics[width=0.8\linewidth]{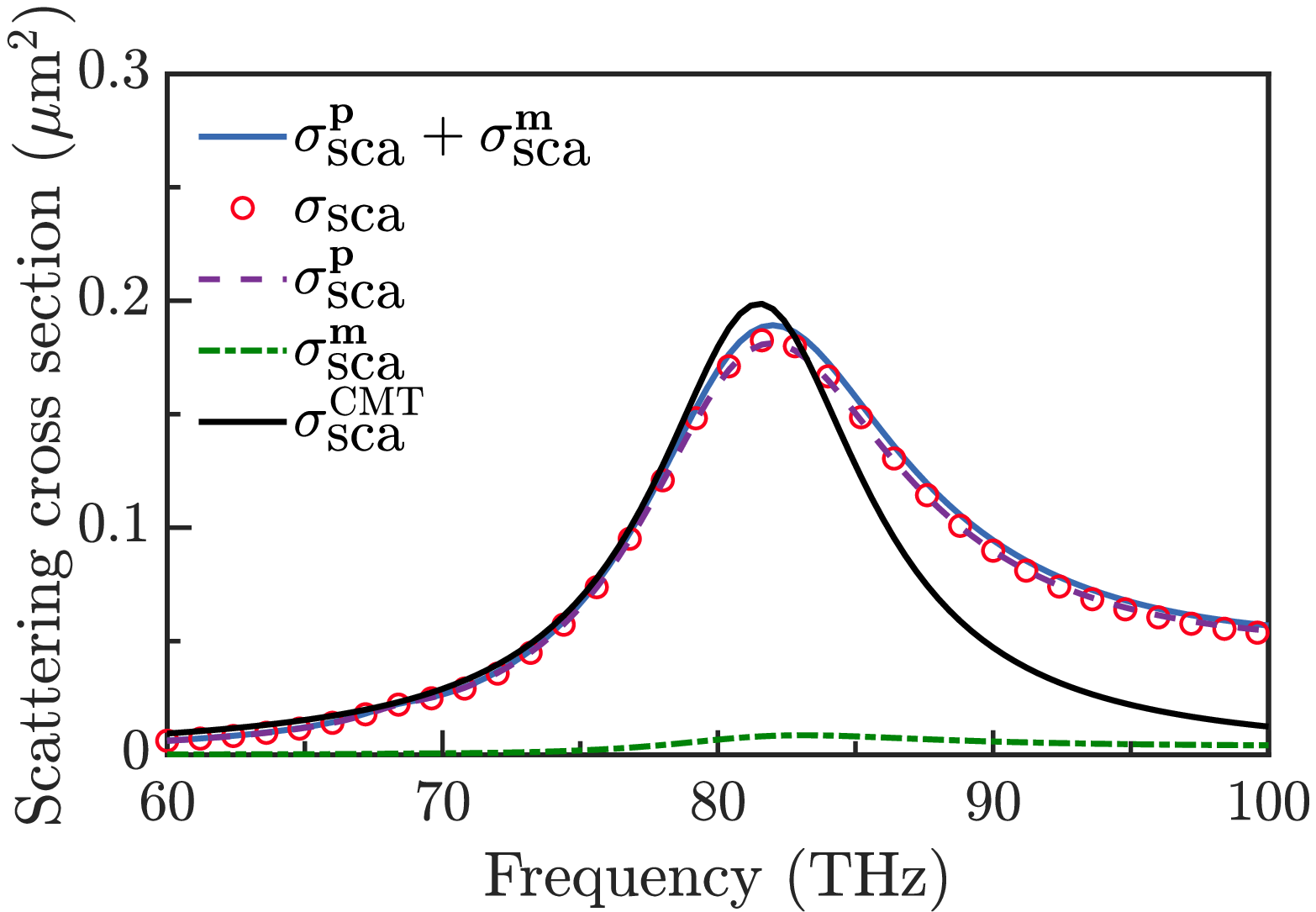}
        \caption{Comparison between analytical and numerical results of scattering cross section in coupled chiral particles of same handedness.}
        \label{fig:8}
    \end{figure}
}

\newcommand{\FigNine}{
    \begin{figure}[tb!]
        \centering
        \includegraphics[width=0.8\linewidth]{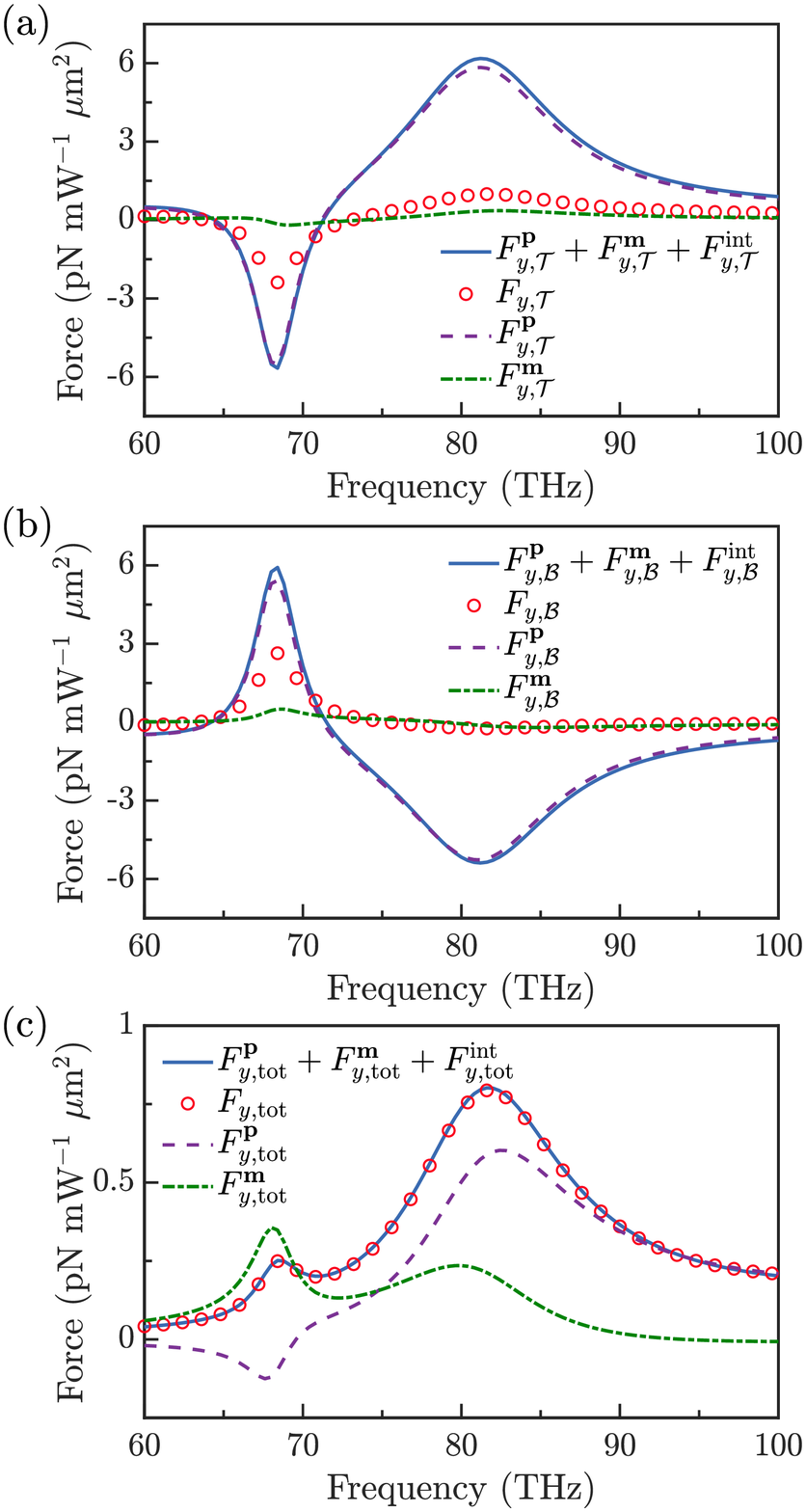}
        \caption{Comparison between the analytical and numerical results of optical force in coupled chiral particles of same handedness. (a) Optical force on the top particle. (b) Optical force on the bottom particle. (c) Total optical force on the whole system.}
        \label{fig:9abc}
    \end{figure}
}

\newcommand{\FigTen}{
    \begin{figure}[tb!]
        \centering
        \includegraphics[width=0.8\linewidth]{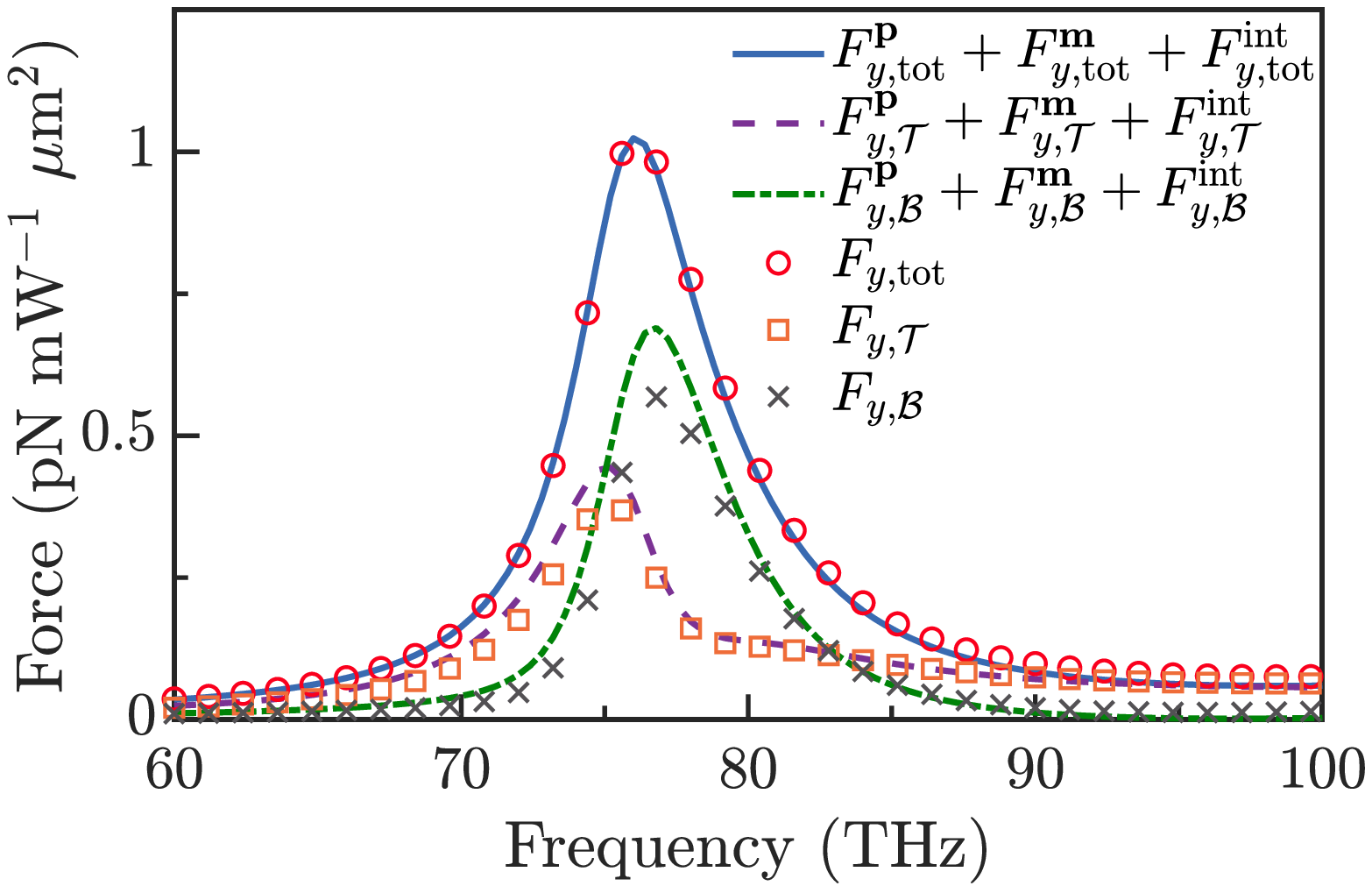}
        \caption{Comparison between the analytical and numerical results of optical force in coupled chiral particles of same handedness with coupling distance $D=1100$ nm.}
        \label{fig:10}
    \end{figure}
}

\newcommand{\FigEleven}{
    \begin{figure}[tb!]
        \centering
        \includegraphics[width=0.8\linewidth]{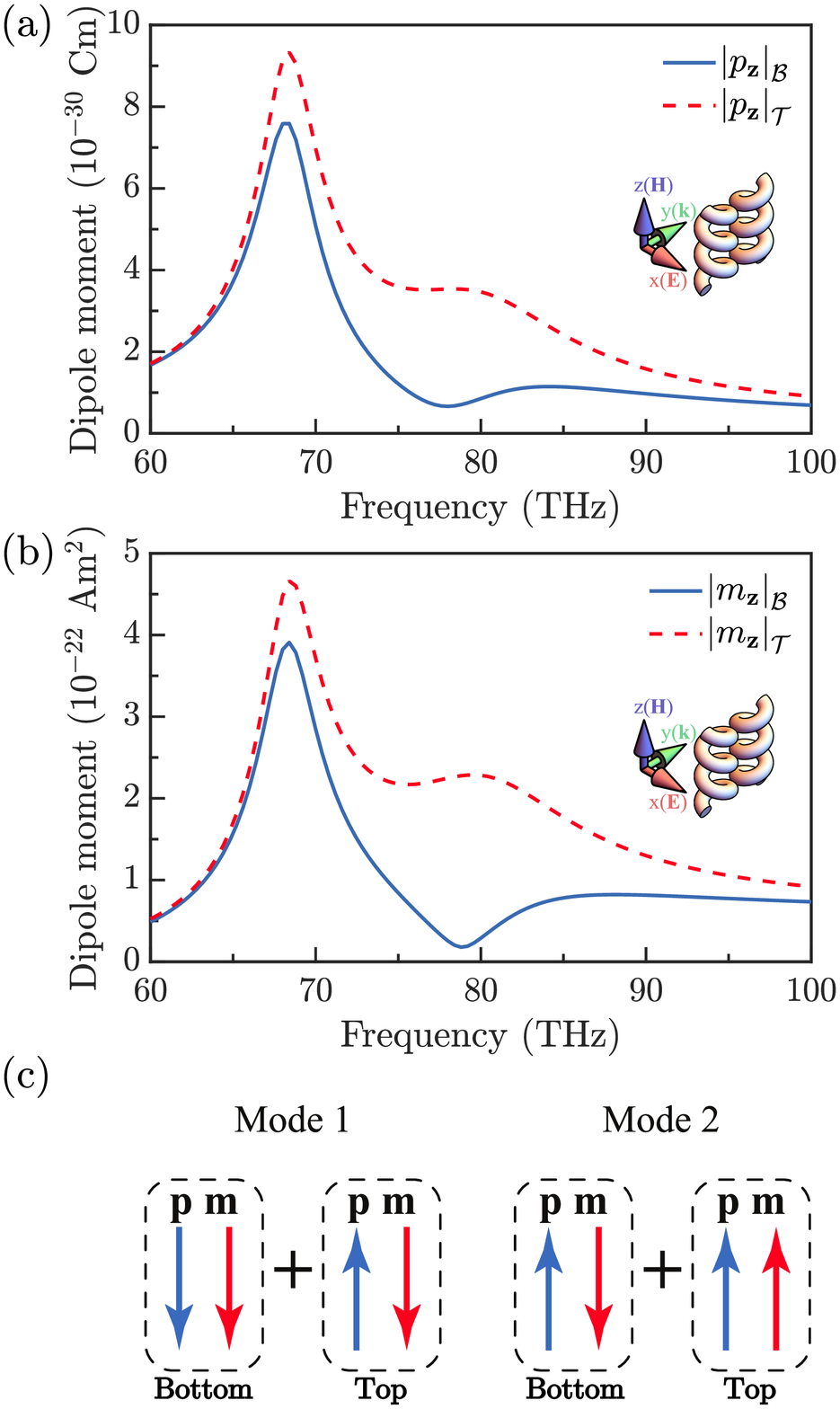}
        \caption{(a) Electric dipole moments and (b) magnetic dipole moments induced in the coupled chiral particles of opposite handedness. (c) Schematics showing the direction of the induced dipole moments at the two resonances.}
        \label{fig:11abc}
    \end{figure}
}

\newcommand{\FigTwelve}{
    \begin{figure}[tb!]
        \centering
        \includegraphics[width=0.8\linewidth]{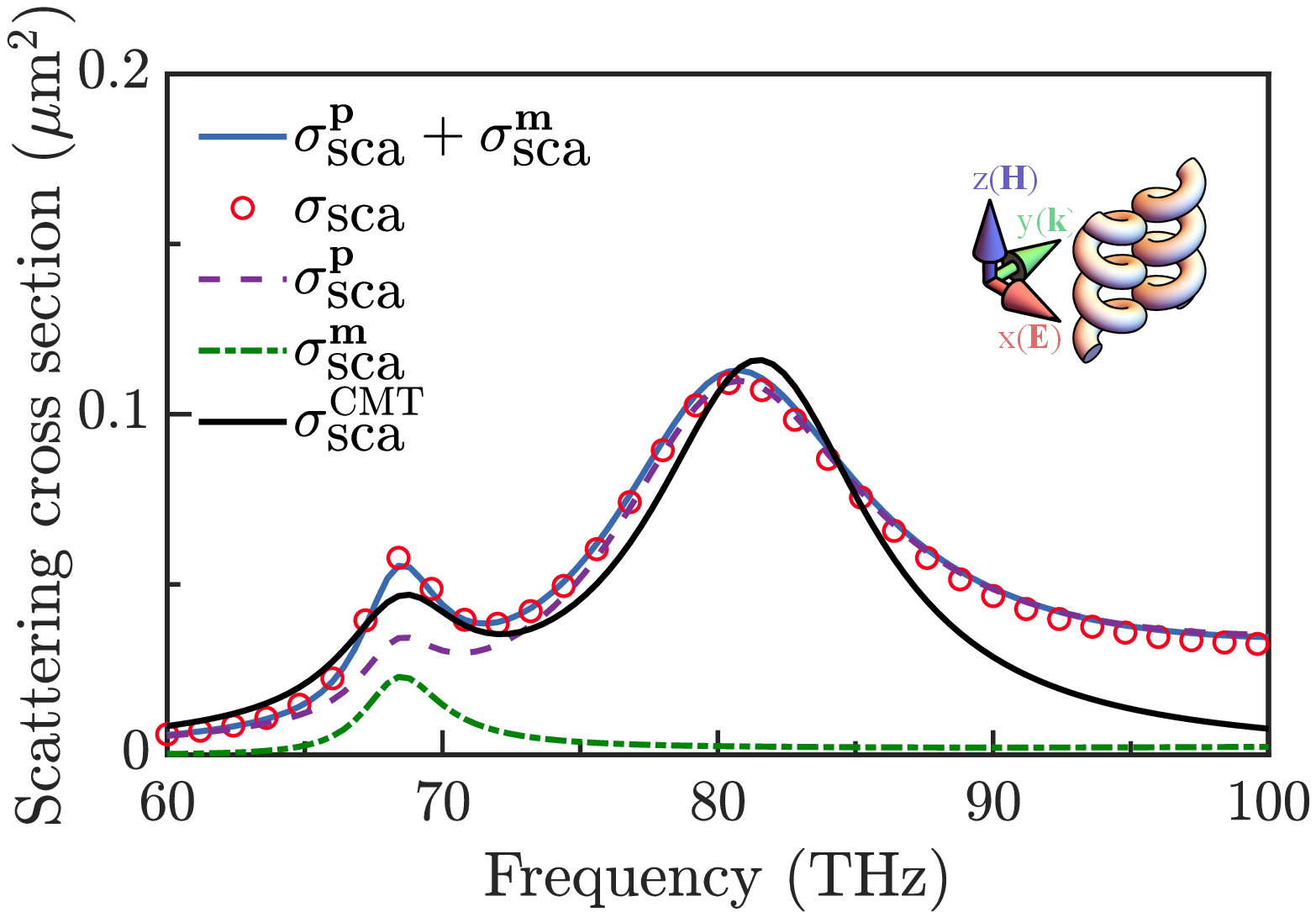}
        \caption{Comparison between analytical and numerical results of scattering cross section in coupled chiral particles of opposite handedness.}
        \label{fig:12}
    \end{figure}
}

\newcommand{\FigThirteen}{
    \begin{figure}[tb!]
        \centering
        \includegraphics[width=0.8\linewidth]{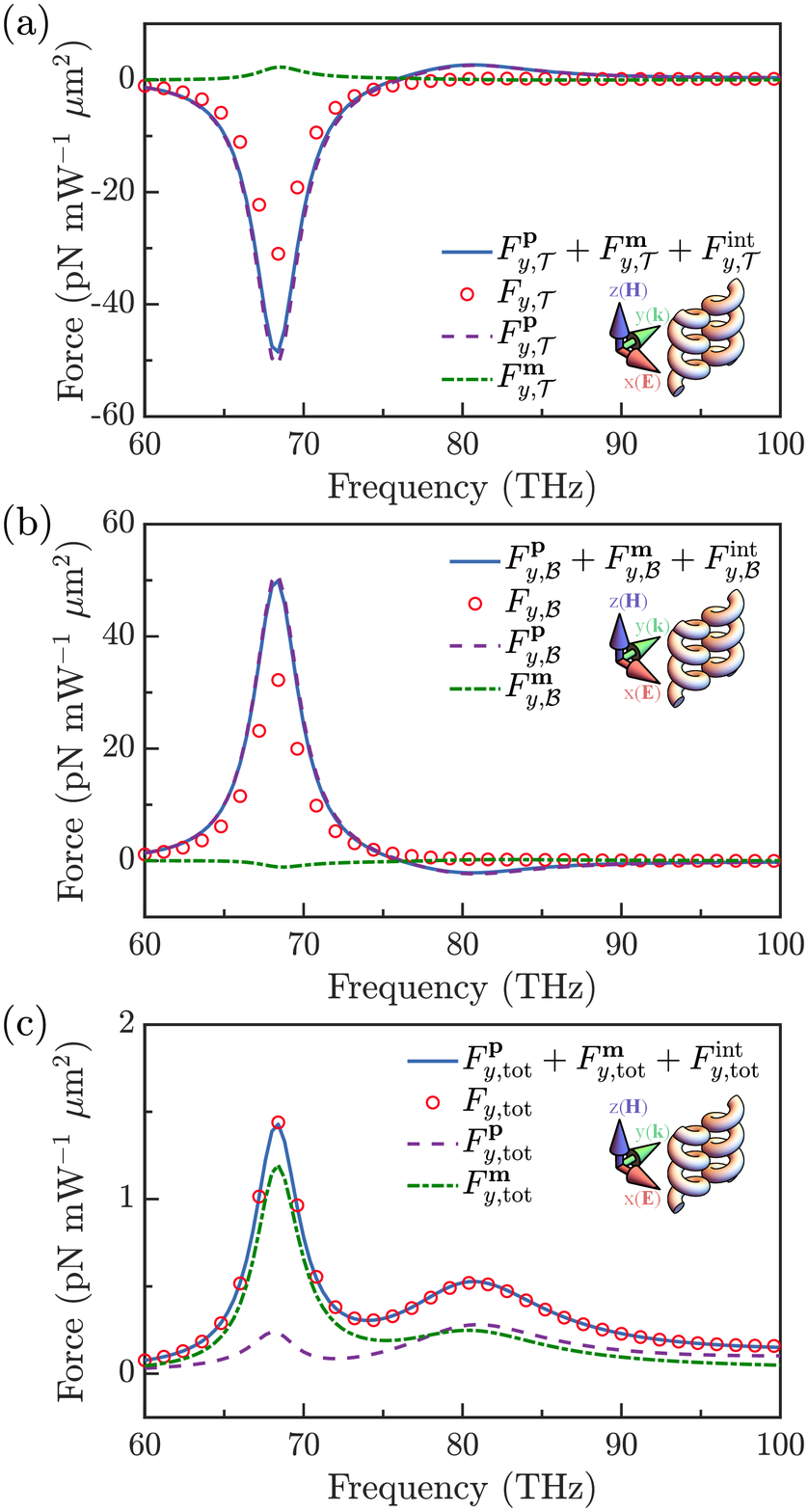}
        \caption{Comparison between analytical and numerical results of optical forces for double helices of opposite handedness. (a) Optical force on the top particle. (b) Optical force on the bottom particle. (c) Total optical force on the whole system.}
        \label{fig:13abc}
    \end{figure}
}

\begin{document}

\title{Optical Forces in Coupled Chiral Particles}
\date{\today}

\author{Kah Jen Wo}
\affiliation{Department of Physics, City University of Hong Kong, Hong Kong, China}
\author{Jie Peng}
\affiliation{Department of Physics, City University of Hong Kong, Hong Kong, China}
\author{Madhava Krishna Prasad}
\affiliation{Department of Physics, City University of Hong Kong, Hong Kong, China}
\author{Yuzhi Shi}
\affiliation{School of Electrical and Electronic Engineering, Nanyang Technological University,
Singapore}
\author{Jensen Li}
\affiliation{Department of Physics, The Hong Kong University of Science and Technology, Hong Kong, China}
\author{Shubo Wang}\email{shubwang@cityu.edu.hk}
\affiliation{Department of Physics, City University of Hong Kong, Hong Kong, China}
\affiliation{City University of Hong Kong Shenzhen Research Institute, Shenzhen, Guangdong, China}

\begin{abstract}
Structural chirality can induce counter-intuitive optical forces due to inherent symmetry properties. While optical forces on a single chiral particle in the Rayleigh regime have been well studied, optical forces in coupled chiral particles remain less explored. By using full-wave numerical simulations and analytical methods of source representation and coupled mode theory, we investigated the optical forces induced by a plane wave on two chiral particles coupling with each other via the evanescent near fields. We found that the induced electric and magnetic dipoles of the chiral particles have complicated couplings that give rise to dark and bright modes. The interaction force between the particles can be either attractive or repulsive, and its magnitude can be significantly enhanced by the resonance modes. The attractive force is much stronger if two particles are of opposite handedness compared with the case of same handedness. The electric dipole force and the magnetic dipole force have the same sign for two particles with the same handedness, while they are of different signs for two particles with opposite handedness. The results can lead to a better understanding of chirality-induced optical forces with potential applications in optical manipulations and chiral light-matter interactions.
\end{abstract}
\maketitle

\section{\label{sec: I. Introduction}Introduction}
Light carries momentum and can apply optical forces to matter through momentum transfer. Optical forces have been widely used to manipulate small particles \cite{Ashkin1970,Grier2003}, molecules \cite{Yang2009}, and biological cells \cite{Ashkin1987}. Recently, there is a growing interest in unusual optical forces such as pulling forces that drag particles towards the light sources  \cite{Chen2011,Novitsky2011,Wang2016} and lateral forces that can separate chiral particles \cite{Wang2014,Hayat2015,Rodriguez-Fortuno2015,Shi2020}. These counterintuitive optical forces can provide novel degrees of freedom to manipulate matter. In contrast to conventional optical forces such as trapping forces which are less sensitive to symmetry properties, unusual optical forces normally arise from special symmetries of light fields or structures. For examples, optical pulling forces can be induced by Bessel beams with conical wavevectors \cite{Chen2011}, and lateral forces can be induced by linearly polarized plane waves on helical particles \cite{Wang2014}. Surprisingly, microscopic lattice symmetries can change photon pressure at an interface to a tension force \cite{Wang2016}. Manipulating these symmetry properties, therefore, can generate rich physics of optical forces which we can explore. 

Chirality breaks mirror symmetry and can give rise to intriguing optical forces through photonic spin-orbit interactions \cite{Wang2014,Bliokh2014,Sukhov2015,Kalhor2016,Alizadeh2016,Chen2018}, where the conversion of light’s spin angular momentum to linear momentum induces asymmetric coupling. Optical forces on chiral particles and chiral structures have been attracting considerable attention not only for the rich physics but also for the potential applications in enantiomer selections \cite{Cameron2014,Tkachenko2014,Bradshaw2014,Chen2014,Bradshaw2015,Zhang2015,Fernandes2015,Canaguier-Durand2015,Bradshaw2015a,Chen2016,Zhao2016,Rahimzadegan2016,Fernandes2016,Zhang2017,Zhao2017,Cameron2017,Kamandi2017,Schnoering2018,Li2019,Kazemi2020}. For a single chiral particle in the Rayleigh regime where dipole approximation can be invoked, the particle can be treated as a combination of electric and magnetic dipoles, and the optical force can be well explained by the interaction between the external fields and the dipoles \cite{Wang2014}. However, it can be complicated if multiple chiral particles couple with each other, in which case the interactions between induced dipoles can dramatically change the scattered fields and hence the optical forces. This scenario is actually more practical as there are usually many chiral particles in enantiomer selections. Understanding the coupling of chiral particles, therefore, is crucial for applications of optical manipulations. In addition, the coupling of chiral particles can generate rich non-Hermitian physics and provide novel degrees of freedom to manipulate light \cite{Wang2019}. We note that optical forces in various coupled non-chiral structures have been investigated, such as coupled ring cavities \cite{Wiederhecker2009}, coupled waveguides \cite{Povinelli2005,Halterman2005, Li2009,Roels2009}, coupled plates \cite{Wang2011,Liu2011a,Zhang2012}, coupled spherical particles \cite{Burns1989,Burns1990, Depasse1994,Chaumet2001,Tatarkova2002,Mohanty2004,Grzegorczyk2006,Guillon2006,Zelenina2007, Rodriguez2008,Dholakia2010,Miljkovic2010,Liu2011,Demergis2012,Frawley2014}, and particle-on-slab \cite{Chaumet2000,Wang2016a}. However, investigation of optical forces in coupled chiral structures remains in its infancy \cite{Chen2015,Shi2020b,Ahsan2020}.

In this paper, we investigated the optical forces induced in coupled chiral particles. The chiral particles have a helical shape and are made of gold whose permittivity is described by a Drude model. We apply full-wave simulations and Maxwell stress tensor to calculate the optical forces acting on each particle and investigate the dependence of the forces on various system parameters including the separation distance, the pitch number, and the handedness. To understand the underlying physics, we employ the source representation and coupled mode theory (CMT) to analytically study the contribution of the induced electric and magnetic dipoles and their mutual interactions. We found that two resonances arise from the coupling of the chiral particles which significantly enhance the interaction force of the two particles. The interaction force can be either attractive or repulsive, and it can be one order of magnitude stronger than the total force of the model system. The electric dipole force always dominates in the interaction force, while the magnetic dipole force can dominate in the total force. In addition, the electric dipole force and the magnetic dipole force have the same sign if the two particles have the same handedness, while they are of different signs if the two particles have opposite handedness.

The paper is organized as follows. In Sec.\ \ref{sec: II. MST,SourceRep,CMT}, we introduce the model system and the methodology. In Sec.\ \ref{sec: III. Results and Discussion}, we  present the numerical results of the scattering cross section and absorption cross section of the model system as well as the numerical results of optical forces. Analytical results based on source representation and CMT will also be presented to explain the phenomena in this section. We then draw the conclusion in Sec.\ \ref{sec: IV. Conclusion}.

\FigOne

\section{\label{sec: II. MST,SourceRep,CMT}Maxwell Stress Tensor Method, Source Representation of Force, and Coupled Mode Theory}
Our model system consists of two chiral particles with a helical shape, as shown in Fig.\ \ref{fig:1}. The helices have inner radii $r=50$ nm, outer radii $R=150$ nm and pitch $p=220$ nm. The axes of the helices are aligned in z direction. The distance between the centers of the two particles is denoted as $D$. We assume the material of the particles is gold with the relative permittivity described by the Drude model $\varepsilon_r=1-\omega_p^2/(\omega^2+i\omega\gamma)$, where the plasmonic frequency is $\omega_p=1.36\times10^{16}$ rad/s and the damping frequency is $\gamma=7.10\times10^{13}$ rad/s \cite{Olmon2012}. The two particles are under the illumination of a plane wave propagating in $y$ direction and linearly polarized in $x$ direction: $\mathbf{E}_\mathrm{inc}=\uvec{x}E_0e^{i\left(ky-\omega t\right)}$. As labelled in Fig.\ \ref{fig:1}, we will refer to the two particles as bottom $\left( \mathcal{B}\right)$ and top $\left(\mathcal{T}\right)$ particles.

To numerically calculate the optical forces acting on the chiral particles, we first conduct full-wave simulations of the model system to obtain the total electromagnetic fields by using COMSOL Multiphysics \cite{COMSOL}. We then evaluate the time-averaged optical force on each particle as
\EqnOne
where
\EqnTwo
is the time-averaged Maxwell stress tensor \cite{JohnDavid1998}. Here ``*'' denotes the complex conjugate, $\delta_{ij}$ is the Kronecker delta, and $S$ is a closed surface on which we carry out the integral. The surface $S$ completely encloses the particle. The total force acting on the whole system can be obtained by adding up the forces of both particles.

In the long wavelength limit, the fields in Eq.\ (\ref{eqn:2}) can be expressed in terms of multipole sources. The lowest orders of the sources are electric dipole $\mathbf{p}$ and magnetic dipole $\mathbf{m}$ whose Cartesian components are given by the following integrals
\EqnThree
where $\mathbf{J}$ is the induced current density on the particles which can be obtained from COMSOL simulations, $\mathbf{r}'$ is the position of the current relative to the center of the particles, and $V$ is the volume of the particles. The time-averaged optical force in Eq.\ (\ref{eqn:1}) for one particle can be expressed as a sum of $\mathbf{F^p}$ , $\mathbf{F^m}$ and $\mathbf{F}^\mathrm{int}$ , corresponding to the forces attributed to the electric dipole and magnetic dipole and their interference, respectively. They can be expressed as \cite{Zhang2015}
\EqnFour
The force due to high order multipoles including toroidal dipole are negligible in our model system. In Eq.\ (\ref{eqn:4}), $\mathbf{E}$ and $\mathbf{B}$ are the total electric and magnetic fields acting on the particle, which can be expressed as $\mathbf{E}=\mathbf{E}_\mathrm{inc}+\mathbf{E}_\mathrm{sca}$ and $\mathbf{B}=\mathbf{B}_\mathrm{inc}+\mathbf{B}_\mathrm{sca}$. Here $\mathbf{E}_\mathrm{sca}$ and $\mathbf{B}_\mathrm{sca}$ are the fields scattered by the other particle which account for the coupling of the two particles. Both electric dipole and magnetic dipole contribute to the total scattered field such that $\mathbf{E}_\mathrm{sca}=\mathbf{E}_\mathrm{sca}^\mathbf{p}+\mathbf{E}_\mathrm{sca}^\mathbf{m}$ and $\mathbf{B}_\mathrm{sca}=\mathbf{B}_\mathrm{sca}^\mathbf{p}+\mathbf{B}_\mathrm{sca}^\mathbf{m}$. These fields can be directly calculated using the following equations once the dipole moments in Eq.\ (\ref{eqn:3}) are obtained \cite{JohnDavid1998}
\EqnFive
where $r=|\mathbf{r}|$ is the radial distance and $\mathbf{n}$ is a unit vector in the direction of $\mathbf{r}$. The Cartesian components of the forces in Eq.\ (\ref{eqn:4}) can be expressed as
\EqnSix
Here Einstein summation convention has been assumed and $\epsilon_{ijk}$ is the Levi-Civita symbol. In this paper, we will focus on the y component of the forces which is the dominating component in the considered configuration of Fig.\ \ref{fig:1}. As the optical forces are intimately related to the scattering properties of the particles, it is helpful to obtain the normalized scattering cross section using the following expressions \cite{JohnDavid1998}
\EqnSeven
where $S_\mathrm{inc}=|E_0|^2/(2Z_0)$ denotes the incident intensity. Equations (\ref{eqn:4})-(\ref{eqn:7}) can be applied to understand the contribution of the electric and magnetic dipoles to the optical forces and scattering properties of the system. However, they cannot provide a intuitive picture to understand the coupling of the dipoles. To this purpose, We apply the temporary coupled mode theory \cite{Fan2003,Suh2004} to our model system. Under the source representation, each particle can be approximated as a combination of an electric dipole and a magnetic dipole. The rate equations for the coupled dipoles of the system can be expressed as
\EqnEightold
where $A_\mathbf{p}^\mathcal{B}$, $A_\mathbf{p}^\mathcal{T}$ ($A_\mathbf{m}^\mathcal{B}$, $A_\mathbf{m}^\mathcal{T}$) are the field amplitudes of the electric (magnetic) dipoles of the bottom and top particles, respectively. $S_\mathrm{TE}^\mathcal{B}$ ($S_\mathrm{TM}^\mathcal{B}$) denotes the amplitude of incident wave with TE (TM) polarization. $\omega_0$ is the eigenfrequency of the electric and magnetic dipoles for a standalone particle. $\Gamma_\mathrm{e}=\left( \gamma+\gamma_\mathrm{e}+\gamma_\mathrm{em}\right)/2$ and $\Gamma_\mathrm{m}=\left( \gamma+\gamma_\mathrm{m}+\gamma_\mathrm{em}\right)/2$ denote the total loss of the electric dipole and the magnetic dipole, respectively.  $\gamma$ denotes the material loss. $\gamma_\mathrm{e}$, $\gamma_\mathrm{m}$ are the loss of the electric and magnetic dipoles, respectively, due to radiation in the same channel (i.e.\ polarized in the same direction as the dipoles). $\gamma_\mathrm{em}$ is the loss of the dipoles due to radiation in the cross-polarized channel. $\kappa_\mathrm{ee}$ and $\kappa_\mathrm{mm}$ denote the couplings of the dipoles. The above equation can be re-written as
\EqnNine
where
\EqnTen

\FigTwo

In the above formulations, we have assumed that the electric dipoles and the magnetic dipoles have no direct coupling due to the orthogonality of their fields and the subwavelength separation of the particles. This is consistent with the fact that electric dipole and magnetic dipole are ``two sides of the same coin'', i.e.\ they belong to a single mode of the helix which can be considered as a damped LC resonance, since the helix can be modelled as a lumped element consisting of a capacitor, an inductor, and a resistor. We have $\kappa_\mathrm{mm}=\kappa_\mathrm{ee}$ for chiral particles of the same handedness and $\kappa_\mathrm{mm}=-\kappa_\mathrm{ee}$ for chial particles of opposite handedness. Define $\kappa_\mathrm{ee}=\kappa$ for simplicity. The eigenfrequencies of the effective Hamiltonian $H$ can be obtained as $\omega_\mathrm{e}=\omega_0\pm\kappa-i\Gamma_\mathrm{e}$ and $\omega_\mathrm{m}=\omega_0\pm\kappa-i\Gamma_\mathrm{m}$. Assume time convention of $e^{-i\omega t}$, Equation (\ref{eqn:8}) or (\ref{eqn:9}) can be solved to obtain the field amplitudes $A_\mathbf{p}^\mathcal{B}$, $A_\mathbf{p}^\mathcal{T}$, $A_\mathbf{m}^\mathcal{B}$, $A_\mathbf{m}^\mathcal{T}$:
\EqnEleven
The total scattering cross section of the whole system can be expressed as  $\left|\left(\sqrt{\gamma_\mathrm{e}}+\sqrt{\gamma_\mathrm{em}}\right)\left( A_\mathbf{p}^\mathcal{B}+A_\mathbf{p}^\mathcal{T}\right)\right|^2+\left|\left(\sqrt{\gamma_\mathrm{m}}+\sqrt{\gamma_\mathrm{em}}\right)\left( A_\mathbf{m}^\mathcal{B}+A_\mathbf{m}^\mathcal{T}\right)\right|^2$. We note that CMT usually can well reproduce the numerical results if the loss of the system is small \cite{Fan2003,Suh2004}. This condition, however, is not fulfilled in our model system due to the material loss of gold and the open boundary conduction, which lead to low quality factors of the modes. Despite this, we will see that the above CMT can qualitatively reproduce the the numerical results and the analytical results given by Eq.\ (\ref{eqn:7}).

\FigThree

\FigFour

\section{\label{sec: III. Results and Discussion}Results and Discussion}
\subsection{\label{subsec: III. A. Scattering and absorption cross sections}Scattering and absorption cross sections}
The optical forces in the model system are determined by both the scattered near fields and far fields. To understand the properties of the scattered fields, we first instigated the scattering cross section and absorption cross section of the model system. The full-wave numerical results of the scattering and absorption cross sections are shown in Fig.\ \ref{fig:2abc}. The distance between the particles is fixed at $D=340$ nm. In Fig.\ \ref{fig:2abc}(a), we notice that two resonances appear in the considered frequency range due to the coupling of the two particles. The first resonance leads to strong absorption but weak scattering. The second resonance leads to strong scattering but relatively weak absorption. Overall, the scattering of the system is stronger than the absorption in the considered frequency range. We will see that these phenomena are attributed to the dark and bright modes of the model system in the presence of the coupling between the induced dipoles (Sec.\ \hyperref[subsec: III. C. Analytical results based on source representation and CMT]{III C}). To understand the effect of pitch number, we then calculated the cross sections from $N=2$ to $N=4$. As shown in Fig.\ \ref{fig:2abc}(a) to (c), the resonance frequencies undergo red-shift, therefore, the particles become more subwavelength and have a smaller scattering cross section. The absorption cross section also reduces as the fields' penetration (i.e.\ skin depth) relative to the wavelength becomes smaller at lower frequencies. We will discuss the physics of these modes in detail in Sec.\ \hyperref[subsec: III. C. Analytical results based on source representation and CMT]{III C}.

The scattering and absorption of the model system are also affected by the coupling of the two chiral particles. To understand this effect, we calculated the cross sections for different coupling distance $D=340 \text{ nm},380 \text{ nm}, 460 \text{ nm}$. As shown in Fig.\ \ref{fig:3ab}, variation of $D$ leads to a more dramatic change of the absorption cross section compared to the change of the scattering cross section. This is expected as the separation of the two chiral particles are subwavelength and their coupling is mainly attributed to the evanescent near fields which determines the absorption of the system. Increasing the separation distance will reduce the coupling strength and leads to a smaller frequency splitting of the two modes, as shown in Fig.\ \ref{fig:3ab}(b). This is consistent with the properties of the total scattering force to be discussed in Sec. \hyperref[subsec: III. B. Numerical results of optical forces]{III B}. We will provide an intuitive explanation of these phenomena using the analytical model in Sec.\ \hyperref[subsec: III. C. Analytical results based on source representation and CMT]{III C}.

\subsection{\label{subsec: III. B. Numerical results of optical forces}Numerical results of optical forces}
The optical force acting on each particle and the total optical force of the model system are calculated using the Maxwell stress tensor approach introduced in Sec.\ \ref{sec: II. MST,SourceRep,CMT}. The results are shown in Figs. \ref{fig:4abcdefghi}-\ref{fig:6abc}. We focus on the optical forces along $y$ direction. Figure \ref{fig:4abcdefghi} (a) shows that the interaction force is attractive at the first resonance and repulsive at the second resonance. In addition, the magnitude of the attractive force is much larger than that of the repulsive force. The total force of the system has two peaks and is always positive (i.e. pushing) since it is a scattering force, as shown in Fig.\ \ref{fig:4abcdefghi}(d). Figure 4(a),(b) and (c) show that, when the pitch number increases from $N=2$ to $N=4$, the magnitude of the interaction forces decrease. This is expected as the resonance frequencies undergoes red-shift as $N$ increases, leading to a smaller dipole moments and thus smaller forces. This is consistent with the properties of the scattering cross section in Fig.\ \ref{fig:2abc}. Similar phenomena also exist for the total forces shown in Fig. \ref{fig:4abcdefghi}(d), (e) and (f). In addition, the frequency splitting of the two peaks increases as the pitch number is increased. The reason is that the coupling due to the evanescent near fields is stronger at lower frequencies. 

\FigFive
\FigSix

The dependence of the optical forces on distance $D$ is also studied, and the results are shown in Figure \ref{fig:5abc}. The magnitudes of the attractive and repulsive forces at the two resonances are enhanced when $D$ decreases from $D=460$ nm to $D=340$ nm, as shown in Fig.\ \ref{fig:5abc}(a) and (b), and meanwhile the frequency splitting of the two resonances are increased. This is expected as the coupling is stronger at smaller distance of $D$. Figure \ref{fig:5abc}(c) shows that the variation of the total force is small except for an enlarged splitting of the two resonance frequencies, which agrees with the results of scattering cross section in Fig.\ \ref{fig:3ab}(a) and confirms the near-field nature of the interaction force and the far-field nature of the total force.

The sign of chirality (i.e. the handedness of the helices) is of critical importance to the optical forces in chiral systems \cite{Wang2014}. To understand the effect of the handedness of the chiral particles, we set $N=3$, $D=340$ nm and calculated the optical forces for two helices with opposite handedness. The model system is shown by the inset in the corner of Fig.\ \ref{fig:6abc}(a). The incident wave remains the same as in Fig.\ \ref{fig:1}. The interaction force acting on individual chiral particle and the total force are shown in Fig.\ \ref{fig:6abc}(a) and (b). We notice that the interaction force at the first resonance is attractive as in the case with the same handedness, but it is one order of magnitude stronger. The interaction force at the second resonance is suppressed. The total force at the first resonance is much stronger than the force at the second resonance. These are attributed to the different orientations of the induced dipole moments compared to that in the case with the same handedness, which is directly determined by the handedness of the helices. We will discuss the physics of these phenomena using a coupled dipole picture in the next section.

\subsection{\label{subsec: III. C. Analytical results based on source representation and CMT}Analytical results based on source representation and CMT}
To understand the above numerical results of cross sections and optical forces, we employ a source representation and treat the chiral particles as electric and magnetic dipoles under long wavelength condition. We use a CMT to understand the coupling of the dipoles which helps to explain the different mode properties at the two resonance frequencies. The interaction forces can be understood as a net result of the Coulomb force and the magnetic force induced by the electric and magnetic dipoles.
 
A standalone chiral particle in Fig.\ \ref{fig:1} has one resonance at $\sim 75$ THz in the considered frequency range with enhanced electric and magnetic dipole moments. When two same chiral particles couple with each other, the degeneracy is broken and two resonances emerge, one at a higher frequency and the other one at a lower frequency. This is confirmed by the dipole moments in Fig.\ \ref{fig:7abc}, which are evaluated using Eq.\ (\ref{eqn:3}). Figure \ref{fig:7abc}(a) and (b) show the amplitudes of the electric dipole moment $p_z$ and magnetic dipole moment $m_z$, respectively, where the blue (red) solid lines denote the dipole moments of the bottom (top) chiral particle. Only the dominating component along $z$ direction (i.e.\ direction of the helix's axis) are considered. The orientations of the dipole moments are schematically shown in Fig.\ \ref{fig:7abc}(c). At the first resonance, the two electric dipoles align in antiparallel directions and interference destructively, lowering the mode's energy. The two magnetic dipole moments also align anti-parallelly and lower the mode’s energy. The combined dipoles, therefore, form an electromagnetic dark mode and suppress the scattering of the system. At the second resonance, the two electric dipole moments align in parallel direction and interference constructively, raising the mode’s energy. The magnetic dipole moments also align parallelly and raise the mode’s energy. The combined dipoles, therefore, form an electromagnetic bright mode and enhance the scattering of the system.

\FigSeven

The scattering cross sections of the dark and bright modes of the coupled chiral particles can be directly determined using Eq.\ (\ref{eqn:7}) after obtaining the dipole moments. The results are shown in Fig.\ \ref{fig:8} for the contributions of electric dipole $\sigma_\mathrm{sca}^\mathbf{p}$ (dashed purple line) and magnetic dipole $\sigma_\mathrm{sca}^\mathbf{m}$ (dashed green line) and $\sigma_\mathrm{sca}^\mathbf{p}+\sigma_\mathrm{sca}^\mathbf{m}$ (solid blue line). As expected, the electric dipole dominates since the particle is of subwavelength, and the scattering cross section is significantly enhanced at the bright mode at $\sim 82$ THz, while it is much smaller at the dark mode at $\sim 68$ THz. The total contribution of $\sigma_\mathrm{sca}^\mathbf{p}+\sigma_\mathrm{sca}^\mathbf{m}$ quantitatively agrees with the full-wave numerical result (symbol line) obtained using COMSOL. To verify the physical picture of coupled dipoles, we fit $\sigma_\mathrm{sca}^\mathbf{p}+\sigma_\mathrm{sca}^\mathbf{m}$ using the expression of scattering cross section predicted by the CMT in Sec.\ \ref{sec: II. MST,SourceRep,CMT}. The fitting result is denoted by the solid black line in Fig.\ \ref{fig:8}, which reasonably well agrees with the analytical result. We notice that the CMT cannot accurately reproduce the analytical result in presence of strong dispersion of gold and the low quality factor of the resonance \cite{Fan2003}.

\FigEight

Using the physical picture of coupled dipoles in Fig.\ \ref{fig:7abc}(c), one can immediately predict the sign of the interaction force between the chiral particles. The force is a sum of the Coulomb force due to the electric dipoles and the magnetic force due to the magnetic dipoles. The Coulomb force dominates as a result of the dominance of electric dipoles. For the electromagnetic dark mode, both the electric dipoles and magnetic dipoles induce an attractive force, i.e.\ a binding force. For the electromagnetic bright mode, both the electric dipoles and magnetic dipoles induce a repulsive force. This agrees with the numerical results in Fig.\ \ref{fig:4abcdefghi}. In addition, we calculated the optical forces using the source representation in Eq.\ (\ref{eqn:6}), and the results are shown in Fig.\ \ref{fig:9abc}. Figure 9(a) and (b) shows the optical forces acting on the top and bottom particles, respectively. We notice that the electric force dominates, as expected, and it is attractive at the first resonance and repulsive at the second resonance. The magnitude of the analytical results (solid blue line) is larger than the numerical result for two reasons: the dipole approximation is not accurate when applied to near-field interactions where evanescent waves with large wavevectors play a dominating role; the size of the chiral particles is comparable to their separation, therefore, neglecting the retardation effect as assumed in the source representation is not appropriate. Despite this, the analytical results can already reproduce the key features of the interaction force, such as the attractive (repulsive) force at the first (second) resonance. The analytical results will approach the full-wave numerical results if the separation of the two chiral particles are larger, in which case the effect of evanescent waves is weak. This is confirmed by the results shown in Fig. 10 for the case of $D=1100$ nm, where the analytical results quantitatively agree with the numerical results for both the individual forces and the total force. Figure \ref{fig:9abc}(c) shows the comparison between the analytical result and the numerical result for the total optical force on the whole system. The analytical result (solid blue line) agrees very well with the numerical result (symbol red line), since the source representation can well describe the far-field properties of the system under long wavelength condition. The electric dipole force and the magnetic dipole force correspond to the addition of the individual dipole forces in Fig. \ref{fig:9abc}(a) and (b). Interestingly, the total electric dipole force is negative at the first resonance.
\FigNine

To understand the effect of chirality on the optical forces, we also apply the source representation and CMT to the case where two chiral particles have opposite handedness. Figure \ref{fig:11abc}(a) and (b) shows the induced electric dipole moment and magnetic dipole moment, respectively. In contrast to the case with the same handedness, the dipole moments are significantly enhanced at the first resonance while are much weaker at the second resonance. The orientations of the dipoles are shown in Fig.\ \ref{fig:11abc}(c). We notice that, at the first resonance, the electric dipoles are anti-parallel and give rise to destructive interference, lowering the energy, while the magnetic dipoles are parallel and give rise to constructive interference, raising the energy. The mode is therefore electrically dark but magnetically bright. This mode can be more easily excited as opposed to the electromagnetic dark mode in the case of Fig.\ \ref{fig:7abc}, which explains the larger dipole moments. At the second resonance, the electric dipoles are parallel and give rise to constructive interference, raising the energy, while the magnetic dipoles are anti-parallel and give rise to destructive interference, lowering the energy. This mode is therefore electrically bright but magnetically dark, and it is more difficult to excite than the electromagnetic bright mode in the case of Fig.\ \ref{fig:7abc}. The above mechanisms derive from the opposite handedness of the chiral particles and gives rise to the enhanced dipole moments at the first resonance and suppressed dipole moments at the second resonance. 
\FigTen
\FigEleven

The dipole moments in Fig.\ \ref{fig:11abc} are used to evaluate the scattering cross sections, and the results are shown in Fig.\ \ref{fig:12}. In contrast to the Fig. 8 for the case of same handedness, at the frist resonance, the contribution of the magnetic dipoles (denoted by the dashed green line) is comparable to that of electric dipoles (denoted by the dashed purple line) due to the magnetic bright mode. The electric dipoles dominate in the second resonance since the corresponding mode is electrically bright but magnetically dark. The total scattering cross section of $\sigma_\mathrm{sca}^\mathbf{p}+\sigma_\mathrm{sca}^\mathbf{m}$ (solid blue line) agrees well with the full-wave numerical result (symbol red line). We fit $\sigma_\mathrm{sca}^\mathbf{p}+\sigma_\mathrm{sca}^\mathbf{m}$ using the expression of scattering cross section predicted by the CMT in Sec.\ \ref{sec: II. MST,SourceRep,CMT}. The fitting result is denoted by the solid black line, which qualitatively agrees with the analytical and numerical results. In addition, we found that the orientations of the dipoles predicted by the CMT agree with the physical picture of source representation in Fig.\ \ref{fig:11abc}(c).

\FigTwelve
\FigThirteen

The optical forces for the coupled chiral particles with opposite handedness can also be understood based on the source representation. As shown in Fig.\ \ref{fig:13abc}(a) and (b), at the first resonance, the attractive force due to the electric dipoles (solid blue line) dominates, the resulting interaction force is one order of magnitude larger than the force in the case with the same handedness, as seen by comparing Fig.\ \ref{fig:9abc}(a),(b) and Fig.\ \ref{fig:13abc}(a),(b). This is due to the larger dipole moments induced at the first resonance. The interaction force at the second resonance is too weak to be observable, in contrast to the case with the same handedness in Fig.\ \ref{fig:9abc}(a) and (b). The reason is twofold: the mode is more difficult to excite than the electromagnetic bright mode in the case with the same handedness; the forces due to electric and magnetic dipoles are in opposite directions. The analytical results of the total interaction force (solid blue lines) qualitatively agree with the numerical results (symbol red lines). The agreement is better than the case of Fig. 9(a) and (b) due to the effective larger separation of the electric dipoles (the ends of the helices, where the charges accumulate, point outwards). The net optical force acting on the two particles is shown in Fig.\ \ref{fig:13abc}(c). The force is entirely attributed to the scattered far fields which can be well described by the source representation. The analytical result (solid blue line), therefore, quantitatively agrees with the numerical result (symbol red line). The electric dipole force and magnetic dipole force correspond to the addition of the individual forces in Fig.\ \ref{fig:13abc}(a) and (b). We notice that the electric dipole force dominates at the first resonance, while at the second resonance it is comparable to the magnetic dipole force.

\section{\label{sec: IV. Conclusion}Conclusion}
In summary, we investigated the optical forces induced by an incident plane wave on two coupled chiral particles. The eigenmodes of the particles in the considered frequency range lead to resonance enhancement of induced electric and magnetic dipoles. The couplings of the dipoles give rise to two new modes, one at a lower frequency and the other at a higher frequency. Using CMT and a source representation, we found that electric dipoles dominate in the near fields and the two modes can be dark or bright depends on the relative orientation of the dipoles. In the case of two particles with the same handedness, the lower frequency mode is an electromagnetic dark mode where both electric and magnetic dipoles align in anti-parallel directions, resulting in suppressed scattering and an attractive force between the particles. The higher frequency mode is an electromagnetic bright mode with both electric and magnetic dipole aligned in parallel direction, resulting in enhanced scattering and a repulsive force. In the case of two particles with opposite handedness, the lower frequency mode is electrically dark but magnetically bright, and the attractive force is significantly enhanced by the resonances of dipoles. In contrast, the higher frequency mode is electrically bright but magnetically dark, where the scattering cross section is dominated by the electric dipoles and the repulsive force is almost negligible.
 
Light-matter interactions associated with chiral structures have been playing critical roles in nanophotonics and metamaterials. The considered model system is probably the simplest structure with near-field couplings fully taken into account, therefore, it can serve as a basic unit to understand the light-matter interactions in periodic chiral structures such as chiral metamaterials and chiral photonic crystals. The results in this paper can also be applied to understand the optical forces in general chiral structures, which could have applications in optical micromanipulations and reconfigurable optical devices. In addition, engineering the couplings in chiral structures can introduce interesting non-Hermitian physics into the systems, where the integration of structural chirality and the chirality of exceptional points (i.e.\ a non-Hermitian degeneracy) can generate novel optical forces.
 
\section{\label{sec: V. Acknowledgements}Acknowledgements}
This work was supported by the National Nature Science Foundation of China (Project No.\ 11904306) and the Research Grants Council of the Hong Kong Special Administrative Region, China (Project Nos.\ HKUST C6013-18G and CityU 11306019).

% \nocite{*}
\bibliography{references_mendeley}% Produces the bibliography via BibTeX.

\begin{thebibliography}{70}%
\makeatletter
\providecommand \@ifxundefined [1]{%
 \@ifx{#1\undefined}
}%
\providecommand \@ifnum [1]{%
 \ifnum #1\expandafter \@firstoftwo
 \else \expandafter \@secondoftwo
 \fi
}%
\providecommand \@ifx [1]{%
 \ifx #1\expandafter \@firstoftwo
 \else \expandafter \@secondoftwo
 \fi
}%
\providecommand \natexlab [1]{#1}%
\providecommand \enquote  [1]{``#1''}%
\providecommand \bibnamefont  [1]{#1}%
\providecommand \bibfnamefont [1]{#1}%
\providecommand \citenamefont [1]{#1}%
\providecommand \href@noop [0]{\@secondoftwo}%
\providecommand \href [0]{\begingroup \@sanitize@url \@href}%
\providecommand \@href[1]{\@@startlink{#1}\@@href}%
\providecommand \@@href[1]{\endgroup#1\@@endlink}%
\providecommand \@sanitize@url [0]{\catcode `\\12\catcode `\$12\catcode
  `\&12\catcode `\#12\catcode `\^12\catcode `\_12\catcode `\%12\relax}%
\providecommand \@@startlink[1]{}%
\providecommand \@@endlink[0]{}%
\providecommand \url  [0]{\begingroup\@sanitize@url \@url }%
\providecommand \@url [1]{\endgroup\@href {#1}{\urlprefix }}%
\providecommand \urlprefix  [0]{URL }%
\providecommand \Eprint [0]{\href }%
\providecommand \doibase [0]{https://doi.org/}%
\providecommand \selectlanguage [0]{\@gobble}%
\providecommand \bibinfo  [0]{\@secondoftwo}%
\providecommand \bibfield  [0]{\@secondoftwo}%
\providecommand \translation [1]{[#1]}%
\providecommand \BibitemOpen [0]{}%
\providecommand \bibitemStop [0]{}%
\providecommand \bibitemNoStop [0]{.\EOS\space}%
\providecommand \EOS [0]{\spacefactor3000\relax}%
\providecommand \BibitemShut  [1]{\csname bibitem#1\endcsname}%
\let\auto@bib@innerbib\@empty
%</preamble>
\bibitem [{\citenamefont {Ashkin}(1970)}]{Ashkin1970}%
  \BibitemOpen
  \bibfield  {author} {\bibinfo {author} {\bibfnamefont {A.}~\bibnamefont
  {Ashkin}},\ }\href {https://doi.org/10.1103/PhysRevLett.24.156} {\bibfield
  {journal} {\bibinfo  {journal} {Phys. Rev. Lett.}\ }\textbf {\bibinfo
  {volume} {24}},\ \bibinfo {pages} {156} (\bibinfo {year} {1970})}\BibitemShut
  {NoStop}%
\bibitem [{\citenamefont {Grier}(2003)}]{Grier2003}%
  \BibitemOpen
  \bibfield  {author} {\bibinfo {author} {\bibfnamefont {D.~G.}\ \bibnamefont
  {Grier}},\ }\href {https://doi.org/10.1038/nature01935} {\bibfield  {journal}
  {\bibinfo  {journal} {Nature}\ }\textbf {\bibinfo {volume} {424}},\ \bibinfo
  {pages} {810} (\bibinfo {year} {2003})}\BibitemShut {NoStop}%
\bibitem [{\citenamefont {Yang}\ \emph {et~al.}(2009)\citenamefont {Yang},
  \citenamefont {Moore}, \citenamefont {Schmidt}, \citenamefont {Klug},
  \citenamefont {Lipson},\ and\ \citenamefont {Erickson}}]{Yang2009}%
  \BibitemOpen
  \bibfield  {author} {\bibinfo {author} {\bibfnamefont {A.~H.~J.}\
  \bibnamefont {Yang}}, \bibinfo {author} {\bibfnamefont {S.~D.}\ \bibnamefont
  {Moore}}, \bibinfo {author} {\bibfnamefont {B.~S.}\ \bibnamefont {Schmidt}},
  \bibinfo {author} {\bibfnamefont {M.}~\bibnamefont {Klug}}, \bibinfo {author}
  {\bibfnamefont {M.}~\bibnamefont {Lipson}},\ and\ \bibinfo {author}
  {\bibfnamefont {D.}~\bibnamefont {Erickson}},\ }\href
  {https://doi.org/10.1038/nature07593} {\bibfield  {journal} {\bibinfo
  {journal} {Nature}\ }\textbf {\bibinfo {volume} {457}},\ \bibinfo {pages}
  {71} (\bibinfo {year} {2009})}\BibitemShut {NoStop}%
\bibitem [{\citenamefont {Ashkin}\ \emph {et~al.}(1987)\citenamefont {Ashkin},
  \citenamefont {Dziedzic},\ and\ \citenamefont {Yamane}}]{Ashkin1987}%
  \BibitemOpen
  \bibfield  {author} {\bibinfo {author} {\bibfnamefont {A.}~\bibnamefont
  {Ashkin}}, \bibinfo {author} {\bibfnamefont {J.~M.}\ \bibnamefont
  {Dziedzic}},\ and\ \bibinfo {author} {\bibfnamefont {T.}~\bibnamefont
  {Yamane}},\ }\href {https://doi.org/10.1038/330769a0} {\bibfield  {journal}
  {\bibinfo  {journal} {Nature}\ }\textbf {\bibinfo {volume} {330}},\ \bibinfo
  {pages} {769} (\bibinfo {year} {1987})}\BibitemShut {NoStop}%
\bibitem [{\citenamefont {Chen}\ \emph {et~al.}(2011)\citenamefont {Chen},
  \citenamefont {Ng}, \citenamefont {Lin},\ and\ \citenamefont
  {Chan}}]{Chen2011}%
  \BibitemOpen
  \bibfield  {author} {\bibinfo {author} {\bibfnamefont {J.}~\bibnamefont
  {Chen}}, \bibinfo {author} {\bibfnamefont {J.}~\bibnamefont {Ng}}, \bibinfo
  {author} {\bibfnamefont {Z.}~\bibnamefont {Lin}},\ and\ \bibinfo {author}
  {\bibfnamefont {C.~T.}\ \bibnamefont {Chan}},\ }\href
  {https://doi.org/10.1038/nphoton.2011.153} {\bibfield  {journal} {\bibinfo
  {journal} {Nat. Photonics}\ }\textbf {\bibinfo {volume} {5}},\ \bibinfo
  {pages} {531} (\bibinfo {year} {2011})}\BibitemShut {NoStop}%
\bibitem [{\citenamefont {Novitsky}\ \emph {et~al.}(2011)\citenamefont
  {Novitsky}, \citenamefont {Qiu},\ and\ \citenamefont {Wang}}]{Novitsky2011}%
  \BibitemOpen
  \bibfield  {author} {\bibinfo {author} {\bibfnamefont {A.}~\bibnamefont
  {Novitsky}}, \bibinfo {author} {\bibfnamefont {C.-W.}\ \bibnamefont {Qiu}},\
  and\ \bibinfo {author} {\bibfnamefont {H.}~\bibnamefont {Wang}},\ }\href
  {https://doi.org/10.1103/PhysRevLett.107.203601} {\bibfield  {journal}
  {\bibinfo  {journal} {Phys. Rev. Lett.}\ }\textbf {\bibinfo {volume} {107}},\
  \bibinfo {pages} {203601} (\bibinfo {year} {2011})}\BibitemShut {NoStop}%
\bibitem [{\citenamefont {Wang}\ \emph {et~al.}(2016)\citenamefont {Wang},
  \citenamefont {Ng}, \citenamefont {Xiao},\ and\ \citenamefont
  {Chan}}]{Wang2016}%
  \BibitemOpen
  \bibfield  {author} {\bibinfo {author} {\bibfnamefont {S.}~\bibnamefont
  {Wang}}, \bibinfo {author} {\bibfnamefont {J.}~\bibnamefont {Ng}}, \bibinfo
  {author} {\bibfnamefont {M.}~\bibnamefont {Xiao}},\ and\ \bibinfo {author}
  {\bibfnamefont {C.~T.}\ \bibnamefont {Chan}},\ }\href
  {https://doi.org/10.1126/sciadv.1501485} {\bibfield  {journal} {\bibinfo
  {journal} {Sci. Adv.}\ }\textbf {\bibinfo {volume} {2}},\ \bibinfo {pages}
  {e1501485} (\bibinfo {year} {2016})}\BibitemShut {NoStop}%
\bibitem [{\citenamefont {Wang}\ and\ \citenamefont {Chan}(2014)}]{Wang2014}%
  \BibitemOpen
  \bibfield  {author} {\bibinfo {author} {\bibfnamefont {S.~B.}\ \bibnamefont
  {Wang}}\ and\ \bibinfo {author} {\bibfnamefont {C.~T.}\ \bibnamefont
  {Chan}},\ }\href {https://doi.org/10.1038/ncomms4307} {\bibfield  {journal}
  {\bibinfo  {journal} {Nat. Commun.}\ }\textbf {\bibinfo {volume} {5}},\
  \bibinfo {pages} {3307} (\bibinfo {year} {2014})}\BibitemShut {NoStop}%
\bibitem [{\citenamefont {Hayat}\ \emph {et~al.}(2015)\citenamefont {Hayat},
  \citenamefont {Mueller},\ and\ \citenamefont {Capasso}}]{Hayat2015}%
  \BibitemOpen
  \bibfield  {author} {\bibinfo {author} {\bibfnamefont {A.}~\bibnamefont
  {Hayat}}, \bibinfo {author} {\bibfnamefont {J.~P.~B.}\ \bibnamefont
  {Mueller}},\ and\ \bibinfo {author} {\bibfnamefont {F.}~\bibnamefont
  {Capasso}},\ }\href {https://doi.org/10.1073/pnas.1516704112} {\bibfield
  {journal} {\bibinfo  {journal} {Proc. Natl. Acad. Sci.}\ }\textbf {\bibinfo
  {volume} {112}},\ \bibinfo {pages} {13190} (\bibinfo {year}
  {2015})}\BibitemShut {NoStop}%
\bibitem [{\citenamefont {Rodr{\'{i}}guez-Fortu{\~{n}}o}\ \emph
  {et~al.}(2015)\citenamefont {Rodr{\'{i}}guez-Fortu{\~{n}}o}, \citenamefont
  {Engheta}, \citenamefont {Mart{\'{i}}nez},\ and\ \citenamefont
  {Zayats}}]{Rodriguez-Fortuno2015}%
  \BibitemOpen
  \bibfield  {author} {\bibinfo {author} {\bibfnamefont {F.~J.}\ \bibnamefont
  {Rodr{\'{i}}guez-Fortu{\~{n}}o}}, \bibinfo {author} {\bibfnamefont
  {N.}~\bibnamefont {Engheta}}, \bibinfo {author} {\bibfnamefont
  {A.}~\bibnamefont {Mart{\'{i}}nez}},\ and\ \bibinfo {author} {\bibfnamefont
  {A.~V.}\ \bibnamefont {Zayats}},\ }\href {https://doi.org/10.1038/ncomms9799}
  {\bibfield  {journal} {\bibinfo  {journal} {Nat. Commun.}\ }\textbf {\bibinfo
  {volume} {6}},\ \bibinfo {pages} {8799} (\bibinfo {year} {2015})}\BibitemShut
  {NoStop}%
\bibitem [{\citenamefont {Shi}\ \emph {et~al.}(2020{\natexlab{a}})\citenamefont
  {Shi}, \citenamefont {Zhu}, \citenamefont {Zhang}, \citenamefont {Mazzulla},
  \citenamefont {Tsai}, \citenamefont {Ding}, \citenamefont {Liu},
  \citenamefont {Cipparrone}, \citenamefont {S{\'{a}}enz},\ and\ \citenamefont
  {Qiu}}]{Shi2020}%
  \BibitemOpen
  \bibfield  {author} {\bibinfo {author} {\bibfnamefont {Y.}~\bibnamefont
  {Shi}}, \bibinfo {author} {\bibfnamefont {T.}~\bibnamefont {Zhu}}, \bibinfo
  {author} {\bibfnamefont {T.}~\bibnamefont {Zhang}}, \bibinfo {author}
  {\bibfnamefont {A.}~\bibnamefont {Mazzulla}}, \bibinfo {author}
  {\bibfnamefont {D.~P.}\ \bibnamefont {Tsai}}, \bibinfo {author}
  {\bibfnamefont {W.}~\bibnamefont {Ding}}, \bibinfo {author} {\bibfnamefont
  {A.~Q.}\ \bibnamefont {Liu}}, \bibinfo {author} {\bibfnamefont
  {G.}~\bibnamefont {Cipparrone}}, \bibinfo {author} {\bibfnamefont {J.~J.}\
  \bibnamefont {S{\'{a}}enz}},\ and\ \bibinfo {author} {\bibfnamefont {C.-W.}\
  \bibnamefont {Qiu}},\ }\href {https://doi.org/10.1038/s41377-020-0293-0}
  {\bibfield  {journal} {\bibinfo  {journal} {Light Sci. Appl.}\ }\textbf
  {\bibinfo {volume} {9}},\ \bibinfo {pages} {62} (\bibinfo {year}
  {2020}{\natexlab{a}})}\BibitemShut {NoStop}%
\bibitem [{\citenamefont {Bliokh}\ \emph {et~al.}(2014)\citenamefont {Bliokh},
  \citenamefont {Bekshaev},\ and\ \citenamefont {Nori}}]{Bliokh2014}%
  \BibitemOpen
  \bibfield  {author} {\bibinfo {author} {\bibfnamefont {K.~Y.}\ \bibnamefont
  {Bliokh}}, \bibinfo {author} {\bibfnamefont {A.~Y.}\ \bibnamefont
  {Bekshaev}},\ and\ \bibinfo {author} {\bibfnamefont {F.}~\bibnamefont
  {Nori}},\ }\href {https://doi.org/10.1038/ncomms4300} {\bibfield  {journal}
  {\bibinfo  {journal} {Nat. Commun.}\ }\textbf {\bibinfo {volume} {5}},\
  \bibinfo {pages} {3300} (\bibinfo {year} {2014})}\BibitemShut {NoStop}%
\bibitem [{\citenamefont {Sukhov}\ \emph {et~al.}(2015)\citenamefont {Sukhov},
  \citenamefont {Kajorndejnukul}, \citenamefont {Naraghi},\ and\ \citenamefont
  {Dogariu}}]{Sukhov2015}%
  \BibitemOpen
  \bibfield  {author} {\bibinfo {author} {\bibfnamefont {S.}~\bibnamefont
  {Sukhov}}, \bibinfo {author} {\bibfnamefont {V.}~\bibnamefont
  {Kajorndejnukul}}, \bibinfo {author} {\bibfnamefont {R.~R.}\ \bibnamefont
  {Naraghi}},\ and\ \bibinfo {author} {\bibfnamefont {A.}~\bibnamefont
  {Dogariu}},\ }\href {https://doi.org/10.1038/nphoton.2015.200} {\bibfield
  {journal} {\bibinfo  {journal} {Nat. Photonics}\ }\textbf {\bibinfo {volume}
  {9}},\ \bibinfo {pages} {809} (\bibinfo {year} {2015})}\BibitemShut {NoStop}%
\bibitem [{\citenamefont {Kalhor}\ \emph {et~al.}(2016)\citenamefont {Kalhor},
  \citenamefont {Thundat},\ and\ \citenamefont {Jacob}}]{Kalhor2016}%
  \BibitemOpen
  \bibfield  {author} {\bibinfo {author} {\bibfnamefont {F.}~\bibnamefont
  {Kalhor}}, \bibinfo {author} {\bibfnamefont {T.}~\bibnamefont {Thundat}},\
  and\ \bibinfo {author} {\bibfnamefont {Z.}~\bibnamefont {Jacob}},\ }\href
  {https://doi.org/10.1063/1.4941539} {\bibfield  {journal} {\bibinfo
  {journal} {Appl. Phys. Lett.}\ }\textbf {\bibinfo {volume} {108}},\ \bibinfo
  {pages} {061102} (\bibinfo {year} {2016})}\BibitemShut {NoStop}%
\bibitem [{\citenamefont {Alizadeh}\ and\ \citenamefont
  {Reinhard}(2016)}]{Alizadeh2016}%
  \BibitemOpen
  \bibfield  {author} {\bibinfo {author} {\bibfnamefont {M.~H.}\ \bibnamefont
  {Alizadeh}}\ and\ \bibinfo {author} {\bibfnamefont {B.~M.}\ \bibnamefont
  {Reinhard}},\ }\href {https://doi.org/10.1364/ol.41.004735} {\bibfield
  {journal} {\bibinfo  {journal} {Opt. Lett.}\ }\textbf {\bibinfo {volume}
  {41}},\ \bibinfo {pages} {4735} (\bibinfo {year} {2016})}\BibitemShut
  {NoStop}%
\bibitem [{\citenamefont {Chen}\ \emph {et~al.}(2018)\citenamefont {Chen},
  \citenamefont {Wang}, \citenamefont {Li},\ and\ \citenamefont
  {Ng}}]{Chen2018}%
  \BibitemOpen
  \bibfield  {author} {\bibinfo {author} {\bibfnamefont {J.}~\bibnamefont
  {Chen}}, \bibinfo {author} {\bibfnamefont {S.}~\bibnamefont {Wang}}, \bibinfo
  {author} {\bibfnamefont {X.}~\bibnamefont {Li}},\ and\ \bibinfo {author}
  {\bibfnamefont {J.}~\bibnamefont {Ng}},\ }\href
  {https://doi.org/10.1364/oe.26.027694} {\bibfield  {journal} {\bibinfo
  {journal} {Opt. Express}\ }\textbf {\bibinfo {volume} {26}},\ \bibinfo
  {pages} {27694} (\bibinfo {year} {2018})}\BibitemShut {NoStop}%
\bibitem [{\citenamefont {Cameron}\ \emph {et~al.}(2014)\citenamefont
  {Cameron}, \citenamefont {Barnett},\ and\ \citenamefont {Yao}}]{Cameron2014}%
  \BibitemOpen
  \bibfield  {author} {\bibinfo {author} {\bibfnamefont {R.~P.}\ \bibnamefont
  {Cameron}}, \bibinfo {author} {\bibfnamefont {S.~M.}\ \bibnamefont
  {Barnett}},\ and\ \bibinfo {author} {\bibfnamefont {A.~M.}\ \bibnamefont
  {Yao}},\ }\href {https://doi.org/10.1088/1367-2630/16/1/013020} {\bibfield
  {journal} {\bibinfo  {journal} {New J. Phys.}\ }\textbf {\bibinfo {volume}
  {16}},\ \bibinfo {pages} {013020} (\bibinfo {year} {2014})}\BibitemShut
  {NoStop}%
\bibitem [{\citenamefont {Tkachenko}\ and\ \citenamefont
  {Brasselet}(2014)}]{Tkachenko2014}%
  \BibitemOpen
  \bibfield  {author} {\bibinfo {author} {\bibfnamefont {G.}~\bibnamefont
  {Tkachenko}}\ and\ \bibinfo {author} {\bibfnamefont {E.}~\bibnamefont
  {Brasselet}},\ }\href {https://doi.org/10.1038/ncomms5491} {\bibfield
  {journal} {\bibinfo  {journal} {Nat. Commun.}\ }\textbf {\bibinfo {volume}
  {5}},\ \bibinfo {pages} {4491} (\bibinfo {year} {2014})}\BibitemShut
  {NoStop}%
\bibitem [{\citenamefont {Bradshaw}\ and\ \citenamefont
  {Andrews}(2014)}]{Bradshaw2014}%
  \BibitemOpen
  \bibfield  {author} {\bibinfo {author} {\bibfnamefont {D.~S.}\ \bibnamefont
  {Bradshaw}}\ and\ \bibinfo {author} {\bibfnamefont {D.~L.}\ \bibnamefont
  {Andrews}},\ }\href {https://doi.org/10.1088/1367-2630/16/10/103021}
  {\bibfield  {journal} {\bibinfo  {journal} {New J. Phys.}\ }\textbf {\bibinfo
  {volume} {16}},\ \bibinfo {pages} {103021} (\bibinfo {year}
  {2014})}\BibitemShut {NoStop}%
\bibitem [{\citenamefont {Chen}\ \emph {et~al.}(2014)\citenamefont {Chen},
  \citenamefont {Wang}, \citenamefont {Lu}, \citenamefont {Liu},\ and\
  \citenamefont {Lin}}]{Chen2014}%
  \BibitemOpen
  \bibfield  {author} {\bibinfo {author} {\bibfnamefont {H.}~\bibnamefont
  {Chen}}, \bibinfo {author} {\bibfnamefont {N.}~\bibnamefont {Wang}}, \bibinfo
  {author} {\bibfnamefont {W.}~\bibnamefont {Lu}}, \bibinfo {author}
  {\bibfnamefont {S.}~\bibnamefont {Liu}},\ and\ \bibinfo {author}
  {\bibfnamefont {Z.}~\bibnamefont {Lin}},\ }\href
  {https://doi.org/10.1103/PhysRevA.90.043850} {\bibfield  {journal} {\bibinfo
  {journal} {Phys. Rev. A}\ }\textbf {\bibinfo {volume} {90}},\ \bibinfo
  {pages} {043850} (\bibinfo {year} {2014})}\BibitemShut {NoStop}%
\bibitem [{\citenamefont {Bradshaw}\ \emph {et~al.}(2015)\citenamefont
  {Bradshaw}, \citenamefont {Forbes}, \citenamefont {Leeder},\ and\
  \citenamefont {Andrews}}]{Bradshaw2015}%
  \BibitemOpen
  \bibfield  {author} {\bibinfo {author} {\bibfnamefont {D.}~\bibnamefont
  {Bradshaw}}, \bibinfo {author} {\bibfnamefont {K.}~\bibnamefont {Forbes}},
  \bibinfo {author} {\bibfnamefont {J.}~\bibnamefont {Leeder}},\ and\ \bibinfo
  {author} {\bibfnamefont {D.}~\bibnamefont {Andrews}},\ }\href
  {https://doi.org/10.3390/photonics2020483} {\bibfield  {journal} {\bibinfo
  {journal} {Photonics}\ }\textbf {\bibinfo {volume} {2}},\ \bibinfo {pages}
  {483} (\bibinfo {year} {2015})}\BibitemShut {NoStop}%
\bibitem [{\citenamefont {Zhang}\ \emph {et~al.}(2015)\citenamefont {Zhang},
  \citenamefont {Wang}, \citenamefont {Lin}, \citenamefont {Sun},\ and\
  \citenamefont {Chan}}]{Zhang2015}%
  \BibitemOpen
  \bibfield  {author} {\bibinfo {author} {\bibfnamefont {X.-L.}\ \bibnamefont
  {Zhang}}, \bibinfo {author} {\bibfnamefont {S.~B.}\ \bibnamefont {Wang}},
  \bibinfo {author} {\bibfnamefont {Z.}~\bibnamefont {Lin}}, \bibinfo {author}
  {\bibfnamefont {H.-B.}\ \bibnamefont {Sun}},\ and\ \bibinfo {author}
  {\bibfnamefont {C.~T.}\ \bibnamefont {Chan}},\ }\href
  {https://doi.org/10.1103/PhysRevA.92.043804} {\bibfield  {journal} {\bibinfo
  {journal} {Phys. Rev. A}\ }\textbf {\bibinfo {volume} {92}},\ \bibinfo
  {pages} {043804} (\bibinfo {year} {2015})}\BibitemShut {NoStop}%
\bibitem [{\citenamefont {Fernandes}\ and\ \citenamefont
  {Silveirinha}(2015)}]{Fernandes2015}%
  \BibitemOpen
  \bibfield  {author} {\bibinfo {author} {\bibfnamefont {D.~E.}\ \bibnamefont
  {Fernandes}}\ and\ \bibinfo {author} {\bibfnamefont {M.~G.}\ \bibnamefont
  {Silveirinha}},\ }\href {https://doi.org/10.1103/PhysRevA.91.061801}
  {\bibfield  {journal} {\bibinfo  {journal} {Phys. Rev. A}\ }\textbf {\bibinfo
  {volume} {91}},\ \bibinfo {pages} {061801} (\bibinfo {year}
  {2015})}\BibitemShut {NoStop}%
\bibitem [{\citenamefont {Canaguier-Durand}\ and\ \citenamefont
  {Genet}(2015)}]{Canaguier-Durand2015}%
  \BibitemOpen
  \bibfield  {author} {\bibinfo {author} {\bibfnamefont {A.}~\bibnamefont
  {Canaguier-Durand}}\ and\ \bibinfo {author} {\bibfnamefont {C.}~\bibnamefont
  {Genet}},\ }\href {https://doi.org/10.1103/PhysRevA.92.043823} {\bibfield
  {journal} {\bibinfo  {journal} {Phys. Rev. A}\ }\textbf {\bibinfo {volume}
  {92}},\ \bibinfo {pages} {043823} (\bibinfo {year} {2015})}\BibitemShut
  {NoStop}%
\bibitem [{\citenamefont {Bradshaw}\ and\ \citenamefont
  {Andrews}(2015)}]{Bradshaw2015a}%
  \BibitemOpen
  \bibfield  {author} {\bibinfo {author} {\bibfnamefont {D.~S.}\ \bibnamefont
  {Bradshaw}}\ and\ \bibinfo {author} {\bibfnamefont {D.~L.}\ \bibnamefont
  {Andrews}},\ }\href {https://doi.org/10.1364/ol.40.000677} {\bibfield
  {journal} {\bibinfo  {journal} {Opt. Lett.}\ }\textbf {\bibinfo {volume}
  {40}},\ \bibinfo {pages} {677} (\bibinfo {year} {2015})}\BibitemShut
  {NoStop}%
\bibitem [{\citenamefont {Chen}\ \emph {et~al.}(2016)\citenamefont {Chen},
  \citenamefont {Liang}, \citenamefont {Liu},\ and\ \citenamefont
  {Lin}}]{Chen2016}%
  \BibitemOpen
  \bibfield  {author} {\bibinfo {author} {\bibfnamefont {H.}~\bibnamefont
  {Chen}}, \bibinfo {author} {\bibfnamefont {C.}~\bibnamefont {Liang}},
  \bibinfo {author} {\bibfnamefont {S.}~\bibnamefont {Liu}},\ and\ \bibinfo
  {author} {\bibfnamefont {Z.}~\bibnamefont {Lin}},\ }\href
  {https://doi.org/10.1103/PhysRevA.93.053833} {\bibfield  {journal} {\bibinfo
  {journal} {Phys. Rev. A}\ }\textbf {\bibinfo {volume} {93}},\ \bibinfo
  {pages} {053833} (\bibinfo {year} {2016})}\BibitemShut {NoStop}%
\bibitem [{\citenamefont {Zhao}\ \emph {et~al.}(2016)\citenamefont {Zhao},
  \citenamefont {Saleh},\ and\ \citenamefont {Dionne}}]{Zhao2016}%
  \BibitemOpen
  \bibfield  {author} {\bibinfo {author} {\bibfnamefont {Y.}~\bibnamefont
  {Zhao}}, \bibinfo {author} {\bibfnamefont {A.~A.~E.}\ \bibnamefont {Saleh}},\
  and\ \bibinfo {author} {\bibfnamefont {J.~A.}\ \bibnamefont {Dionne}},\
  }\href {https://doi.org/10.1021/acsphotonics.5b00574} {\bibfield  {journal}
  {\bibinfo  {journal} {ACS Photonics}\ }\textbf {\bibinfo {volume} {3}},\
  \bibinfo {pages} {304} (\bibinfo {year} {2016})}\BibitemShut {NoStop}%
\bibitem [{\citenamefont {Rahimzadegan}\ \emph {et~al.}(2016)\citenamefont
  {Rahimzadegan}, \citenamefont {Fruhnert}, \citenamefont {Alaee},
  \citenamefont {Fernandez-Corbaton},\ and\ \citenamefont
  {Rockstuhl}}]{Rahimzadegan2016}%
  \BibitemOpen
  \bibfield  {author} {\bibinfo {author} {\bibfnamefont {A.}~\bibnamefont
  {Rahimzadegan}}, \bibinfo {author} {\bibfnamefont {M.}~\bibnamefont
  {Fruhnert}}, \bibinfo {author} {\bibfnamefont {R.}~\bibnamefont {Alaee}},
  \bibinfo {author} {\bibfnamefont {I.}~\bibnamefont {Fernandez-Corbaton}},\
  and\ \bibinfo {author} {\bibfnamefont {C.}~\bibnamefont {Rockstuhl}},\ }\href
  {https://doi.org/10.1103/PhysRevB.94.125123} {\bibfield  {journal} {\bibinfo
  {journal} {Phys. Rev. B}\ }\textbf {\bibinfo {volume} {94}},\ \bibinfo
  {pages} {125123} (\bibinfo {year} {2016})}\BibitemShut {NoStop}%
\bibitem [{\citenamefont {Fernandes}\ and\ \citenamefont
  {Silveirinha}(2016)}]{Fernandes2016}%
  \BibitemOpen
  \bibfield  {author} {\bibinfo {author} {\bibfnamefont {D.~E.}\ \bibnamefont
  {Fernandes}}\ and\ \bibinfo {author} {\bibfnamefont {M.~G.}\ \bibnamefont
  {Silveirinha}},\ }\href {https://doi.org/10.1103/PhysRevApplied.6.014016}
  {\bibfield  {journal} {\bibinfo  {journal} {Phys. Rev. Appl.}\ }\textbf
  {\bibinfo {volume} {6}},\ \bibinfo {pages} {014016} (\bibinfo {year}
  {2016})}\BibitemShut {NoStop}%
\bibitem [{\citenamefont {Zhang}\ \emph {et~al.}(2017)\citenamefont {Zhang},
  \citenamefont {Mahdy}, \citenamefont {Liu}, \citenamefont {Teng},
  \citenamefont {Lim}, \citenamefont {Wang},\ and\ \citenamefont
  {Qiu}}]{Zhang2017}%
  \BibitemOpen
  \bibfield  {author} {\bibinfo {author} {\bibfnamefont {T.}~\bibnamefont
  {Zhang}}, \bibinfo {author} {\bibfnamefont {M.~R.~C.}\ \bibnamefont {Mahdy}},
  \bibinfo {author} {\bibfnamefont {Y.}~\bibnamefont {Liu}}, \bibinfo {author}
  {\bibfnamefont {J.~H.}\ \bibnamefont {Teng}}, \bibinfo {author}
  {\bibfnamefont {C.~T.}\ \bibnamefont {Lim}}, \bibinfo {author} {\bibfnamefont
  {Z.}~\bibnamefont {Wang}},\ and\ \bibinfo {author} {\bibfnamefont {C.-W.}\
  \bibnamefont {Qiu}},\ }\href {https://doi.org/10.1021/acsnano.7b01428}
  {\bibfield  {journal} {\bibinfo  {journal} {ACS Nano}\ }\textbf {\bibinfo
  {volume} {11}},\ \bibinfo {pages} {4292} (\bibinfo {year}
  {2017})}\BibitemShut {NoStop}%
\bibitem [{\citenamefont {Zhao}\ \emph {et~al.}(2017)\citenamefont {Zhao},
  \citenamefont {Saleh}, \citenamefont {van~de Haar}, \citenamefont {Baum},
  \citenamefont {Briggs}, \citenamefont {Lay}, \citenamefont {Reyes-Becerra},\
  and\ \citenamefont {Dionne}}]{Zhao2017}%
  \BibitemOpen
  \bibfield  {author} {\bibinfo {author} {\bibfnamefont {Y.}~\bibnamefont
  {Zhao}}, \bibinfo {author} {\bibfnamefont {A.~A.~E.}\ \bibnamefont {Saleh}},
  \bibinfo {author} {\bibfnamefont {M.~A.}\ \bibnamefont {van~de Haar}},
  \bibinfo {author} {\bibfnamefont {B.}~\bibnamefont {Baum}}, \bibinfo {author}
  {\bibfnamefont {J.~A.}\ \bibnamefont {Briggs}}, \bibinfo {author}
  {\bibfnamefont {A.}~\bibnamefont {Lay}}, \bibinfo {author} {\bibfnamefont
  {O.~A.}\ \bibnamefont {Reyes-Becerra}},\ and\ \bibinfo {author}
  {\bibfnamefont {J.~A.}\ \bibnamefont {Dionne}},\ }\href
  {https://doi.org/10.1038/nnano.2017.180} {\bibfield  {journal} {\bibinfo
  {journal} {Nat. Nanotechnol.}\ }\textbf {\bibinfo {volume} {12}},\ \bibinfo
  {pages} {1055} (\bibinfo {year} {2017})}\BibitemShut {NoStop}%
\bibitem [{\citenamefont {Cameron}\ \emph {et~al.}(2017)\citenamefont
  {Cameron}, \citenamefont {G{\"{o}}tte}, \citenamefont {Barnett},\ and\
  \citenamefont {Yao}}]{Cameron2017}%
  \BibitemOpen
  \bibfield  {author} {\bibinfo {author} {\bibfnamefont {R.~P.}\ \bibnamefont
  {Cameron}}, \bibinfo {author} {\bibfnamefont {J.~B.}\ \bibnamefont
  {G{\"{o}}tte}}, \bibinfo {author} {\bibfnamefont {S.~M.}\ \bibnamefont
  {Barnett}},\ and\ \bibinfo {author} {\bibfnamefont {A.~M.}\ \bibnamefont
  {Yao}},\ }\href {https://doi.org/10.1098/rsta.2015.0433} {\bibfield
  {journal} {\bibinfo  {journal} {Philos. Trans. R. Soc. A Math. Phys. Eng.
  Sci.}\ }\textbf {\bibinfo {volume} {375}},\ \bibinfo {pages} {20150433}
  (\bibinfo {year} {2017})}\BibitemShut {NoStop}%
\bibitem [{\citenamefont {Kamandi}\ \emph {et~al.}(2017)\citenamefont
  {Kamandi}, \citenamefont {Albooyeh}, \citenamefont {Guclu}, \citenamefont
  {Veysi}, \citenamefont {Zeng}, \citenamefont {Wickramasinghe},\ and\
  \citenamefont {Capolino}}]{Kamandi2017}%
  \BibitemOpen
  \bibfield  {author} {\bibinfo {author} {\bibfnamefont {M.}~\bibnamefont
  {Kamandi}}, \bibinfo {author} {\bibfnamefont {M.}~\bibnamefont {Albooyeh}},
  \bibinfo {author} {\bibfnamefont {C.}~\bibnamefont {Guclu}}, \bibinfo
  {author} {\bibfnamefont {M.}~\bibnamefont {Veysi}}, \bibinfo {author}
  {\bibfnamefont {J.}~\bibnamefont {Zeng}}, \bibinfo {author} {\bibfnamefont
  {K.}~\bibnamefont {Wickramasinghe}},\ and\ \bibinfo {author} {\bibfnamefont
  {F.}~\bibnamefont {Capolino}},\ }\href
  {https://doi.org/10.1103/PhysRevApplied.8.064010} {\bibfield  {journal}
  {\bibinfo  {journal} {Phys. Rev. Appl.}\ }\textbf {\bibinfo {volume} {8}},\
  \bibinfo {pages} {064010} (\bibinfo {year} {2017})}\BibitemShut {NoStop}%
\bibitem [{\citenamefont {Schnoering}\ \emph {et~al.}(2018)\citenamefont
  {Schnoering}, \citenamefont {Poulikakos}, \citenamefont {Rosales-Cabara},
  \citenamefont {Canaguier-Durand}, \citenamefont {Norris},\ and\ \citenamefont
  {Genet}}]{Schnoering2018}%
  \BibitemOpen
  \bibfield  {author} {\bibinfo {author} {\bibfnamefont {G.}~\bibnamefont
  {Schnoering}}, \bibinfo {author} {\bibfnamefont {L.~V.}\ \bibnamefont
  {Poulikakos}}, \bibinfo {author} {\bibfnamefont {Y.}~\bibnamefont
  {Rosales-Cabara}}, \bibinfo {author} {\bibfnamefont {A.}~\bibnamefont
  {Canaguier-Durand}}, \bibinfo {author} {\bibfnamefont {D.~J.}\ \bibnamefont
  {Norris}},\ and\ \bibinfo {author} {\bibfnamefont {C.}~\bibnamefont
  {Genet}},\ }\href {https://doi.org/10.1103/PhysRevLett.121.023902} {\bibfield
   {journal} {\bibinfo  {journal} {Phys. Rev. Lett.}\ }\textbf {\bibinfo
  {volume} {121}},\ \bibinfo {pages} {023902} (\bibinfo {year}
  {2018})}\BibitemShut {NoStop}%
\bibitem [{\citenamefont {Li}\ \emph {et~al.}(2019)\citenamefont {Li},
  \citenamefont {Yan}, \citenamefont {Zhang}, \citenamefont {Liang},
  \citenamefont {Zhang},\ and\ \citenamefont {Yao}}]{Li2019}%
  \BibitemOpen
  \bibfield  {author} {\bibinfo {author} {\bibfnamefont {M.}~\bibnamefont
  {Li}}, \bibinfo {author} {\bibfnamefont {S.}~\bibnamefont {Yan}}, \bibinfo
  {author} {\bibfnamefont {Y.}~\bibnamefont {Zhang}}, \bibinfo {author}
  {\bibfnamefont {Y.}~\bibnamefont {Liang}}, \bibinfo {author} {\bibfnamefont
  {P.}~\bibnamefont {Zhang}},\ and\ \bibinfo {author} {\bibfnamefont
  {B.}~\bibnamefont {Yao}},\ }\href
  {https://doi.org/10.1103/PhysRevA.99.033825} {\bibfield  {journal} {\bibinfo
  {journal} {Phys. Rev. A}\ }\textbf {\bibinfo {volume} {99}},\ \bibinfo
  {pages} {033825} (\bibinfo {year} {2019})}\BibitemShut {NoStop}%
\bibitem [{\citenamefont {Kazemi}\ and\ \citenamefont
  {Mahmoudi}(2020)}]{Kazemi2020}%
  \BibitemOpen
  \bibfield  {author} {\bibinfo {author} {\bibfnamefont {S.~H.}\ \bibnamefont
  {Kazemi}}\ and\ \bibinfo {author} {\bibfnamefont {M.}~\bibnamefont
  {Mahmoudi}},\ }\href {https://doi.org/10.1088/1402-4896/ab57a1} {\bibfield
  {journal} {\bibinfo  {journal} {Phys. Scr.}\ }\textbf {\bibinfo {volume}
  {95}},\ \bibinfo {pages} {035405} (\bibinfo {year} {2020})}\BibitemShut
  {NoStop}%
\bibitem [{\citenamefont {Wang}\ \emph {et~al.}(2019)\citenamefont {Wang},
  \citenamefont {Hou}, \citenamefont {Lu}, \citenamefont {Chen}, \citenamefont
  {Zhang},\ and\ \citenamefont {Chan}}]{Wang2019}%
  \BibitemOpen
  \bibfield  {author} {\bibinfo {author} {\bibfnamefont {S.}~\bibnamefont
  {Wang}}, \bibinfo {author} {\bibfnamefont {B.}~\bibnamefont {Hou}}, \bibinfo
  {author} {\bibfnamefont {W.}~\bibnamefont {Lu}}, \bibinfo {author}
  {\bibfnamefont {Y.}~\bibnamefont {Chen}}, \bibinfo {author} {\bibfnamefont
  {Z.~Q.}\ \bibnamefont {Zhang}},\ and\ \bibinfo {author} {\bibfnamefont
  {C.~T.}\ \bibnamefont {Chan}},\ }\href
  {https://doi.org/10.1038/s41467-019-08826-6} {\bibfield  {journal} {\bibinfo
  {journal} {Nat. Commun.}\ }\textbf {\bibinfo {volume} {10}},\ \bibinfo
  {pages} {832} (\bibinfo {year} {2019})}\BibitemShut {NoStop}%
\bibitem [{\citenamefont {Wiederhecker}\ \emph {et~al.}(2009)\citenamefont
  {Wiederhecker}, \citenamefont {Chen}, \citenamefont {Gondarenko},\ and\
  \citenamefont {Lipson}}]{Wiederhecker2009}%
  \BibitemOpen
  \bibfield  {author} {\bibinfo {author} {\bibfnamefont {G.~S.}\ \bibnamefont
  {Wiederhecker}}, \bibinfo {author} {\bibfnamefont {L.}~\bibnamefont {Chen}},
  \bibinfo {author} {\bibfnamefont {A.}~\bibnamefont {Gondarenko}},\ and\
  \bibinfo {author} {\bibfnamefont {M.}~\bibnamefont {Lipson}},\ }\href
  {https://doi.org/10.1038/nature08584} {\bibfield  {journal} {\bibinfo
  {journal} {Nature}\ }\textbf {\bibinfo {volume} {462}},\ \bibinfo {pages}
  {633} (\bibinfo {year} {2009})}\BibitemShut {NoStop}%
\bibitem [{\citenamefont {Povinelli}\ \emph {et~al.}(2005)\citenamefont
  {Povinelli}, \citenamefont {Lon{\v{c}}ar}, \citenamefont {Ibanescu},
  \citenamefont {Smythe}, \citenamefont {Johnson}, \citenamefont {Capasso},\
  and\ \citenamefont {Joannopoulos}}]{Povinelli2005}%
  \BibitemOpen
  \bibfield  {author} {\bibinfo {author} {\bibfnamefont {M.~L.}\ \bibnamefont
  {Povinelli}}, \bibinfo {author} {\bibfnamefont {M.}~\bibnamefont
  {Lon{\v{c}}ar}}, \bibinfo {author} {\bibfnamefont {M.}~\bibnamefont
  {Ibanescu}}, \bibinfo {author} {\bibfnamefont {E.~J.}\ \bibnamefont
  {Smythe}}, \bibinfo {author} {\bibfnamefont {S.~G.}\ \bibnamefont {Johnson}},
  \bibinfo {author} {\bibfnamefont {F.}~\bibnamefont {Capasso}},\ and\ \bibinfo
  {author} {\bibfnamefont {J.~D.}\ \bibnamefont {Joannopoulos}},\ }\href
  {https://doi.org/10.1364/ol.30.003042} {\bibfield  {journal} {\bibinfo
  {journal} {Opt. Lett.}\ }\textbf {\bibinfo {volume} {30}},\ \bibinfo {pages}
  {3042} (\bibinfo {year} {2005})}\BibitemShut {NoStop}%
\bibitem [{\citenamefont {Halterman}\ \emph {et~al.}(2005)\citenamefont
  {Halterman}, \citenamefont {Elson},\ and\ \citenamefont
  {Singh}}]{Halterman2005}%
  \BibitemOpen
  \bibfield  {author} {\bibinfo {author} {\bibfnamefont {K.}~\bibnamefont
  {Halterman}}, \bibinfo {author} {\bibfnamefont {J.~M.}\ \bibnamefont
  {Elson}},\ and\ \bibinfo {author} {\bibfnamefont {S.}~\bibnamefont {Singh}},\
  }\href {https://doi.org/10.1103/PhysRevB.72.075429} {\bibfield  {journal}
  {\bibinfo  {journal} {Phys. Rev. B}\ }\textbf {\bibinfo {volume} {72}},\
  \bibinfo {pages} {075429} (\bibinfo {year} {2005})}\BibitemShut {NoStop}%
\bibitem [{\citenamefont {Li}\ \emph {et~al.}(2009)\citenamefont {Li},
  \citenamefont {Pernice},\ and\ \citenamefont {Tang}}]{Li2009}%
  \BibitemOpen
  \bibfield  {author} {\bibinfo {author} {\bibfnamefont {M.}~\bibnamefont
  {Li}}, \bibinfo {author} {\bibfnamefont {W.~H.}\ \bibnamefont {Pernice}},\
  and\ \bibinfo {author} {\bibfnamefont {H.~X.}\ \bibnamefont {Tang}},\ }\href
  {https://doi.org/10.1038/nphoton.2009.116} {\bibfield  {journal} {\bibinfo
  {journal} {Nat. Photonics}\ }\textbf {\bibinfo {volume} {3}},\ \bibinfo
  {pages} {464} (\bibinfo {year} {2009})}\BibitemShut {NoStop}%
\bibitem [{\citenamefont {Roels}\ \emph {et~al.}(2009)\citenamefont {Roels},
  \citenamefont {{De Vlaminck}}, \citenamefont {Lagae}, \citenamefont {Maes},
  \citenamefont {{Van Thourhout}},\ and\ \citenamefont {Baets}}]{Roels2009}%
  \BibitemOpen
  \bibfield  {author} {\bibinfo {author} {\bibfnamefont {J.}~\bibnamefont
  {Roels}}, \bibinfo {author} {\bibfnamefont {I.}~\bibnamefont {{De
  Vlaminck}}}, \bibinfo {author} {\bibfnamefont {L.}~\bibnamefont {Lagae}},
  \bibinfo {author} {\bibfnamefont {B.}~\bibnamefont {Maes}}, \bibinfo {author}
  {\bibfnamefont {D.}~\bibnamefont {{Van Thourhout}}},\ and\ \bibinfo {author}
  {\bibfnamefont {R.}~\bibnamefont {Baets}},\ }\href
  {https://doi.org/10.1038/nnano.2009.186} {\bibfield  {journal} {\bibinfo
  {journal} {Nat. Nanotechnol.}\ }\textbf {\bibinfo {volume} {4}},\ \bibinfo
  {pages} {510} (\bibinfo {year} {2009})}\BibitemShut {NoStop}%
\bibitem [{\citenamefont {Wang}\ \emph {et~al.}(2011)\citenamefont {Wang},
  \citenamefont {Ng}, \citenamefont {Liu}, \citenamefont {Zheng}, \citenamefont
  {Hang},\ and\ \citenamefont {Chan}}]{Wang2011}%
  \BibitemOpen
  \bibfield  {author} {\bibinfo {author} {\bibfnamefont {S.~B.}\ \bibnamefont
  {Wang}}, \bibinfo {author} {\bibfnamefont {J.}~\bibnamefont {Ng}}, \bibinfo
  {author} {\bibfnamefont {H.}~\bibnamefont {Liu}}, \bibinfo {author}
  {\bibfnamefont {H.~H.}\ \bibnamefont {Zheng}}, \bibinfo {author}
  {\bibfnamefont {Z.~H.}\ \bibnamefont {Hang}},\ and\ \bibinfo {author}
  {\bibfnamefont {C.~T.}\ \bibnamefont {Chan}},\ }\href
  {https://doi.org/10.1103/PhysRevB.84.075114} {\bibfield  {journal} {\bibinfo
  {journal} {Phys. Rev. B}\ }\textbf {\bibinfo {volume} {84}},\ \bibinfo
  {pages} {075114} (\bibinfo {year} {2011})}\BibitemShut {NoStop}%
\bibitem [{\citenamefont {Liu}\ \emph {et~al.}(2011{\natexlab{a}})\citenamefont
  {Liu}, \citenamefont {Ng}, \citenamefont {Wang}, \citenamefont {Lin},
  \citenamefont {Hang}, \citenamefont {Chan},\ and\ \citenamefont
  {Zhu}}]{Liu2011a}%
  \BibitemOpen
  \bibfield  {author} {\bibinfo {author} {\bibfnamefont {H.}~\bibnamefont
  {Liu}}, \bibinfo {author} {\bibfnamefont {J.}~\bibnamefont {Ng}}, \bibinfo
  {author} {\bibfnamefont {S.~B.}\ \bibnamefont {Wang}}, \bibinfo {author}
  {\bibfnamefont {Z.~F.}\ \bibnamefont {Lin}}, \bibinfo {author} {\bibfnamefont
  {Z.~H.}\ \bibnamefont {Hang}}, \bibinfo {author} {\bibfnamefont {C.~T.}\
  \bibnamefont {Chan}},\ and\ \bibinfo {author} {\bibfnamefont {S.~N.}\
  \bibnamefont {Zhu}},\ }\href {https://doi.org/10.1103/PhysRevLett.106.087401}
  {\bibfield  {journal} {\bibinfo  {journal} {Phys. Rev. Lett.}\ }\textbf
  {\bibinfo {volume} {106}},\ \bibinfo {pages} {087401} (\bibinfo {year}
  {2011}{\natexlab{a}})}\BibitemShut {NoStop}%
\bibitem [{\citenamefont {Zhang}\ \emph {et~al.}(2012)\citenamefont {Zhang},
  \citenamefont {MacDonald},\ and\ \citenamefont {Zheludev}}]{Zhang2012}%
  \BibitemOpen
  \bibfield  {author} {\bibinfo {author} {\bibfnamefont {J.}~\bibnamefont
  {Zhang}}, \bibinfo {author} {\bibfnamefont {K.~F.}\ \bibnamefont
  {MacDonald}},\ and\ \bibinfo {author} {\bibfnamefont {N.~I.}\ \bibnamefont
  {Zheludev}},\ }\href {https://doi.org/10.1103/PhysRevB.85.205123} {\bibfield
  {journal} {\bibinfo  {journal} {Phys. Rev. B}\ }\textbf {\bibinfo {volume}
  {85}},\ \bibinfo {pages} {205123} (\bibinfo {year} {2012})}\BibitemShut
  {NoStop}%
\bibitem [{\citenamefont {Burns}\ \emph {et~al.}(1989)\citenamefont {Burns},
  \citenamefont {Fournier},\ and\ \citenamefont {Golovchenko}}]{Burns1989}%
  \BibitemOpen
  \bibfield  {author} {\bibinfo {author} {\bibfnamefont {M.~M.}\ \bibnamefont
  {Burns}}, \bibinfo {author} {\bibfnamefont {J.-M.}\ \bibnamefont
  {Fournier}},\ and\ \bibinfo {author} {\bibfnamefont {J.~A.}\ \bibnamefont
  {Golovchenko}},\ }\href {https://doi.org/10.1103/PhysRevLett.63.1233}
  {\bibfield  {journal} {\bibinfo  {journal} {Phys. Rev. Lett.}\ }\textbf
  {\bibinfo {volume} {63}},\ \bibinfo {pages} {1233} (\bibinfo {year}
  {1989})}\BibitemShut {NoStop}%
\bibitem [{\citenamefont {Burns}\ \emph {et~al.}(1990)\citenamefont {Burns},
  \citenamefont {Fournier},\ and\ \citenamefont {Golovchenko}}]{Burns1990}%
  \BibitemOpen
  \bibfield  {author} {\bibinfo {author} {\bibfnamefont {M.~M.}\ \bibnamefont
  {Burns}}, \bibinfo {author} {\bibfnamefont {J.-M.}\ \bibnamefont
  {Fournier}},\ and\ \bibinfo {author} {\bibfnamefont {J.~A.}\ \bibnamefont
  {Golovchenko}},\ }\href {https://doi.org/10.1126/science.249.4970.749}
  {\bibfield  {journal} {\bibinfo  {journal} {Science}\ }\textbf {\bibinfo
  {volume} {249}},\ \bibinfo {pages} {749} (\bibinfo {year}
  {1990})}\BibitemShut {NoStop}%
\bibitem [{\citenamefont {Dapasse}\ and\ \citenamefont
  {Vigoureux}(1994)}]{Depasse1994}%
  \BibitemOpen
  \bibfield  {author} {\bibinfo {author} {\bibfnamefont {F.}~\bibnamefont
  {Dapasse}}\ and\ \bibinfo {author} {\bibfnamefont {J.~M.}\ \bibnamefont
  {Vigoureux}},\ }\href {https://doi.org/10.1088/0022-3727/27/5/006} {\bibfield
   {journal} {\bibinfo  {journal} {J. Phys. D. Appl. Phys.}\ }\textbf {\bibinfo
  {volume} {27}},\ \bibinfo {pages} {914} (\bibinfo {year} {1994})}\BibitemShut
  {NoStop}%
\bibitem [{\citenamefont {Chaumet}\ and\ \citenamefont
  {Nieto-Vesperinas}(2001)}]{Chaumet2001}%
  \BibitemOpen
  \bibfield  {author} {\bibinfo {author} {\bibfnamefont {P.~C.}\ \bibnamefont
  {Chaumet}}\ and\ \bibinfo {author} {\bibfnamefont {M.}~\bibnamefont
  {Nieto-Vesperinas}},\ }\href {https://doi.org/10.1103/PhysRevB.64.035422}
  {\bibfield  {journal} {\bibinfo  {journal} {Phys. Rev. B}\ }\textbf {\bibinfo
  {volume} {64}},\ \bibinfo {pages} {035422} (\bibinfo {year}
  {2001})}\BibitemShut {NoStop}%
\bibitem [{\citenamefont {Tatarkova}\ \emph {et~al.}(2002)\citenamefont
  {Tatarkova}, \citenamefont {Carruthers},\ and\ \citenamefont
  {Dholakia}}]{Tatarkova2002}%
  \BibitemOpen
  \bibfield  {author} {\bibinfo {author} {\bibfnamefont {S.~A.}\ \bibnamefont
  {Tatarkova}}, \bibinfo {author} {\bibfnamefont {A.~E.}\ \bibnamefont
  {Carruthers}},\ and\ \bibinfo {author} {\bibfnamefont {K.}~\bibnamefont
  {Dholakia}},\ }\href {https://doi.org/10.1103/PhysRevLett.89.283901}
  {\bibfield  {journal} {\bibinfo  {journal} {Phys. Rev. Lett.}\ }\textbf
  {\bibinfo {volume} {89}},\ \bibinfo {pages} {283901} (\bibinfo {year}
  {2002})}\BibitemShut {NoStop}%
\bibitem [{\citenamefont {Mohanty}\ \emph {et~al.}(2004)\citenamefont
  {Mohanty}, \citenamefont {Andrews},\ and\ \citenamefont
  {Gupta}}]{Mohanty2004}%
  \BibitemOpen
  \bibfield  {author} {\bibinfo {author} {\bibfnamefont {S.~K.}\ \bibnamefont
  {Mohanty}}, \bibinfo {author} {\bibfnamefont {J.~T.}\ \bibnamefont
  {Andrews}},\ and\ \bibinfo {author} {\bibfnamefont {P.~K.}\ \bibnamefont
  {Gupta}},\ }\href {https://doi.org/10.1364/opex.12.002746} {\bibfield
  {journal} {\bibinfo  {journal} {Opt. Express}\ }\textbf {\bibinfo {volume}
  {12}},\ \bibinfo {pages} {2746} (\bibinfo {year} {2004})}\BibitemShut
  {NoStop}%
\bibitem [{\citenamefont {Grzegorczyk}\ \emph {et~al.}(2006)\citenamefont
  {Grzegorczyk}, \citenamefont {Kemp},\ and\ \citenamefont
  {Kong}}]{Grzegorczyk2006}%
  \BibitemOpen
  \bibfield  {author} {\bibinfo {author} {\bibfnamefont {T.~M.}\ \bibnamefont
  {Grzegorczyk}}, \bibinfo {author} {\bibfnamefont {B.~A.}\ \bibnamefont
  {Kemp}},\ and\ \bibinfo {author} {\bibfnamefont {J.~A.}\ \bibnamefont
  {Kong}},\ }\href {https://doi.org/10.1103/PhysRevLett.96.113903} {\bibfield
  {journal} {\bibinfo  {journal} {Phys. Rev. Lett.}\ }\textbf {\bibinfo
  {volume} {96}},\ \bibinfo {pages} {113903} (\bibinfo {year}
  {2006})}\BibitemShut {NoStop}%
\bibitem [{\citenamefont {Guillon}\ \emph {et~al.}(2006)\citenamefont
  {Guillon}, \citenamefont {Moine},\ and\ \citenamefont {Stout}}]{Guillon2006}%
  \BibitemOpen
  \bibfield  {author} {\bibinfo {author} {\bibfnamefont {M.}~\bibnamefont
  {Guillon}}, \bibinfo {author} {\bibfnamefont {O.}~\bibnamefont {Moine}},\
  and\ \bibinfo {author} {\bibfnamefont {B.}~\bibnamefont {Stout}},\ }\href
  {https://doi.org/10.1103/PhysRevLett.96.143902} {\bibfield  {journal}
  {\bibinfo  {journal} {Phys. Rev. Lett.}\ }\textbf {\bibinfo {volume} {96}},\
  \bibinfo {pages} {143902} (\bibinfo {year} {2006})}\BibitemShut {NoStop}%
\bibitem [{\citenamefont {Zelenina}\ \emph {et~al.}(2007)\citenamefont
  {Zelenina}, \citenamefont {Quidant},\ and\ \citenamefont
  {Nieto-Vesperinas}}]{Zelenina2007}%
  \BibitemOpen
  \bibfield  {author} {\bibinfo {author} {\bibfnamefont {A.~S.}\ \bibnamefont
  {Zelenina}}, \bibinfo {author} {\bibfnamefont {R.}~\bibnamefont {Quidant}},\
  and\ \bibinfo {author} {\bibfnamefont {M.}~\bibnamefont {Nieto-Vesperinas}},\
  }\href {https://doi.org/10.1364/ol.32.001156} {\bibfield  {journal} {\bibinfo
   {journal} {Opt. Lett.}\ }\textbf {\bibinfo {volume} {32}},\ \bibinfo {pages}
  {1156} (\bibinfo {year} {2007})}\BibitemShut {NoStop}%
\bibitem [{\citenamefont {Rodr{\'{i}}guez}\ \emph {et~al.}(2008)\citenamefont
  {Rodr{\'{i}}guez}, \citenamefont {{D{\'{a}}vila Romero}},\ and\ \citenamefont
  {Andrews}}]{Rodriguez2008}%
  \BibitemOpen
  \bibfield  {author} {\bibinfo {author} {\bibfnamefont {J.}~\bibnamefont
  {Rodr{\'{i}}guez}}, \bibinfo {author} {\bibfnamefont {L.~C.}\ \bibnamefont
  {{D{\'{a}}vila Romero}}},\ and\ \bibinfo {author} {\bibfnamefont {D.~L.}\
  \bibnamefont {Andrews}},\ }\href {https://doi.org/10.1103/PhysRevA.78.043805}
  {\bibfield  {journal} {\bibinfo  {journal} {Phys. Rev. A}\ }\textbf {\bibinfo
  {volume} {78}},\ \bibinfo {pages} {043805} (\bibinfo {year}
  {2008})}\BibitemShut {NoStop}%
\bibitem [{\citenamefont {Dholakia}\ and\ \citenamefont
  {Zem{\'{a}}nek}(2010)}]{Dholakia2010}%
  \BibitemOpen
  \bibfield  {author} {\bibinfo {author} {\bibfnamefont {K.}~\bibnamefont
  {Dholakia}}\ and\ \bibinfo {author} {\bibfnamefont {P.}~\bibnamefont
  {Zem{\'{a}}nek}},\ }\href {https://doi.org/10.1103/RevModPhys.82.1767}
  {\bibfield  {journal} {\bibinfo  {journal} {Rev. Mod. Phys.}\ }\textbf
  {\bibinfo {volume} {82}},\ \bibinfo {pages} {1767} (\bibinfo {year}
  {2010})}\BibitemShut {NoStop}%
\bibitem [{\citenamefont {Miljkovi{\'{c}}}\ \emph {et~al.}(2010)\citenamefont
  {Miljkovi{\'{c}}}, \citenamefont {Pakizeh}, \citenamefont {Sepulveda},
  \citenamefont {Johansson},\ and\ \citenamefont {K{\"{a}}ll}}]{Miljkovic2010}%
  \BibitemOpen
  \bibfield  {author} {\bibinfo {author} {\bibfnamefont {V.~D.}\ \bibnamefont
  {Miljkovi{\'{c}}}}, \bibinfo {author} {\bibfnamefont {T.}~\bibnamefont
  {Pakizeh}}, \bibinfo {author} {\bibfnamefont {B.}~\bibnamefont {Sepulveda}},
  \bibinfo {author} {\bibfnamefont {P.}~\bibnamefont {Johansson}},\ and\
  \bibinfo {author} {\bibfnamefont {M.}~\bibnamefont {K{\"{a}}ll}},\ }\href
  {https://doi.org/10.1021/jp911371r} {\bibfield  {journal} {\bibinfo
  {journal} {J. Phys. Chem. C}\ }\textbf {\bibinfo {volume} {114}},\ \bibinfo
  {pages} {7472} (\bibinfo {year} {2010})}\BibitemShut {NoStop}%
\bibitem [{\citenamefont {Liu}\ \emph {et~al.}(2011{\natexlab{b}})\citenamefont
  {Liu}, \citenamefont {Ng}, \citenamefont {Wang}, \citenamefont {Hang},
  \citenamefont {Chan},\ and\ \citenamefont {Zhu}}]{Liu2011}%
  \BibitemOpen
  \bibfield  {author} {\bibinfo {author} {\bibfnamefont {H.}~\bibnamefont
  {Liu}}, \bibinfo {author} {\bibfnamefont {J.}~\bibnamefont {Ng}}, \bibinfo
  {author} {\bibfnamefont {S.~B.}\ \bibnamefont {Wang}}, \bibinfo {author}
  {\bibfnamefont {Z.~H.}\ \bibnamefont {Hang}}, \bibinfo {author}
  {\bibfnamefont {C.~T.}\ \bibnamefont {Chan}},\ and\ \bibinfo {author}
  {\bibfnamefont {S.~N.}\ \bibnamefont {Zhu}},\ }\href
  {https://doi.org/10.1088/1367-2630/13/7/073040} {\bibfield  {journal}
  {\bibinfo  {journal} {New J. Phys.}\ }\textbf {\bibinfo {volume} {13}},\
  \bibinfo {pages} {073040} (\bibinfo {year} {2011}{\natexlab{b}})}\BibitemShut
  {NoStop}%
\bibitem [{\citenamefont {Demergis}\ and\ \citenamefont
  {Florin}(2012)}]{Demergis2012}%
  \BibitemOpen
  \bibfield  {author} {\bibinfo {author} {\bibfnamefont {V.}~\bibnamefont
  {Demergis}}\ and\ \bibinfo {author} {\bibfnamefont {E.-L.}\ \bibnamefont
  {Florin}},\ }\href {https://doi.org/10.1021/nl303035p} {\bibfield  {journal}
  {\bibinfo  {journal} {Nano Lett.}\ }\textbf {\bibinfo {volume} {12}},\
  \bibinfo {pages} {5756} (\bibinfo {year} {2012})}\BibitemShut {NoStop}%
\bibitem [{\citenamefont {Frawley}\ \emph {et~al.}(2014)\citenamefont
  {Frawley}, \citenamefont {Gusachenko}, \citenamefont {Truong}, \citenamefont
  {Sergides},\ and\ \citenamefont {Chormaic}}]{Frawley2014}%
  \BibitemOpen
  \bibfield  {author} {\bibinfo {author} {\bibfnamefont {M.~C.}\ \bibnamefont
  {Frawley}}, \bibinfo {author} {\bibfnamefont {I.}~\bibnamefont {Gusachenko}},
  \bibinfo {author} {\bibfnamefont {V.~G.}\ \bibnamefont {Truong}}, \bibinfo
  {author} {\bibfnamefont {M.}~\bibnamefont {Sergides}},\ and\ \bibinfo
  {author} {\bibfnamefont {S.~N.}\ \bibnamefont {Chormaic}},\ }\href
  {https://doi.org/10.1364/oe.22.016322} {\bibfield  {journal} {\bibinfo
  {journal} {Opt. Express}\ }\textbf {\bibinfo {volume} {22}},\ \bibinfo
  {pages} {16322} (\bibinfo {year} {2014})}\BibitemShut {NoStop}%
\bibitem [{\citenamefont {Chaumet}\ and\ \citenamefont
  {Nieto-Vesperinas}(2000)}]{Chaumet2000}%
  \BibitemOpen
  \bibfield  {author} {\bibinfo {author} {\bibfnamefont {P.~C.}\ \bibnamefont
  {Chaumet}}\ and\ \bibinfo {author} {\bibfnamefont {M.}~\bibnamefont
  {Nieto-Vesperinas}},\ }\href {https://doi.org/10.1103/PhysRevB.61.14119}
  {\bibfield  {journal} {\bibinfo  {journal} {Phys. Rev. B}\ }\textbf {\bibinfo
  {volume} {61}},\ \bibinfo {pages} {14119} (\bibinfo {year}
  {2000})}\BibitemShut {NoStop}%
\bibitem [{\citenamefont {Wang}\ and\ \citenamefont {Chan}(2016)}]{Wang2016a}%
  \BibitemOpen
  \bibfield  {author} {\bibinfo {author} {\bibfnamefont {S.}~\bibnamefont
  {Wang}}\ and\ \bibinfo {author} {\bibfnamefont {C.~T.}\ \bibnamefont
  {Chan}},\ }\href {https://doi.org/10.1364/oe.24.002235} {\bibfield  {journal}
  {\bibinfo  {journal} {Opt. Express}\ }\textbf {\bibinfo {volume} {24}},\
  \bibinfo {pages} {2235} (\bibinfo {year} {2016})}\BibitemShut {NoStop}%
\bibitem [{\citenamefont {Chen}\ \emph {et~al.}(2015)\citenamefont {Chen},
  \citenamefont {Jiang}, \citenamefont {Wang}, \citenamefont {Lu},
  \citenamefont {Liu},\ and\ \citenamefont {Lin}}]{Chen2015}%
  \BibitemOpen
  \bibfield  {author} {\bibinfo {author} {\bibfnamefont {H.}~\bibnamefont
  {Chen}}, \bibinfo {author} {\bibfnamefont {Y.}~\bibnamefont {Jiang}},
  \bibinfo {author} {\bibfnamefont {N.}~\bibnamefont {Wang}}, \bibinfo {author}
  {\bibfnamefont {W.}~\bibnamefont {Lu}}, \bibinfo {author} {\bibfnamefont
  {S.}~\bibnamefont {Liu}},\ and\ \bibinfo {author} {\bibfnamefont
  {Z.}~\bibnamefont {Lin}},\ }\href {https://doi.org/10.1364/ol.40.005530}
  {\bibfield  {journal} {\bibinfo  {journal} {Opt. Lett.}\ }\textbf {\bibinfo
  {volume} {40}},\ \bibinfo {pages} {5530} (\bibinfo {year}
  {2015})}\BibitemShut {NoStop}%
\bibitem [{\citenamefont {Shi}\ \emph {et~al.}(2020{\natexlab{b}})\citenamefont
  {Shi}, \citenamefont {Zheng}, \citenamefont {Chen}, \citenamefont {Lu},
  \citenamefont {Liu},\ and\ \citenamefont {Lin}}]{Shi2020b}%
  \BibitemOpen
  \bibfield  {author} {\bibinfo {author} {\bibfnamefont {H.}~\bibnamefont
  {Shi}}, \bibinfo {author} {\bibfnamefont {H.}~\bibnamefont {Zheng}}, \bibinfo
  {author} {\bibfnamefont {H.}~\bibnamefont {Chen}}, \bibinfo {author}
  {\bibfnamefont {W.}~\bibnamefont {Lu}}, \bibinfo {author} {\bibfnamefont
  {S.}~\bibnamefont {Liu}},\ and\ \bibinfo {author} {\bibfnamefont
  {Z.}~\bibnamefont {Lin}},\ }\href
  {https://doi.org/10.1103/PhysRevA.101.043808} {\bibfield  {journal} {\bibinfo
   {journal} {Phys. Rev. A}\ }\textbf {\bibinfo {volume} {101}},\ \bibinfo
  {pages} {043808} (\bibinfo {year} {2020}{\natexlab{b}})}\BibitemShut
  {NoStop}%
\bibitem [{\citenamefont {Ahsan}\ \emph {et~al.}(2020)\citenamefont {Ahsan},
  \citenamefont {Shamim}, \citenamefont {Mahdy}, \citenamefont {Das},
  \citenamefont {Rivy}, \citenamefont {Dolon}, \citenamefont {Hossain},\ and\
  \citenamefont {Faisal}}]{Ahsan2020}%
  \BibitemOpen
  \bibfield  {author} {\bibinfo {author} {\bibfnamefont {N.~B.}\ \bibnamefont
  {Ahsan}}, \bibinfo {author} {\bibfnamefont {R.}~\bibnamefont {Shamim}},
  \bibinfo {author} {\bibfnamefont {M.~R.~C.}\ \bibnamefont {Mahdy}}, \bibinfo
  {author} {\bibfnamefont {S.~C.}\ \bibnamefont {Das}}, \bibinfo {author}
  {\bibfnamefont {H.~M.}\ \bibnamefont {Rivy}}, \bibinfo {author}
  {\bibfnamefont {C.~I.}\ \bibnamefont {Dolon}}, \bibinfo {author}
  {\bibfnamefont {M.}~\bibnamefont {Hossain}},\ and\ \bibinfo {author}
  {\bibfnamefont {K.~M.}\ \bibnamefont {Faisal}},\ }\href
  {https://doi.org/10.1364/JOSAB.383004} {\bibfield  {journal} {\bibinfo
  {journal} {J. Opt. Soc. Am. B}\ }\textbf {\bibinfo {volume} {37}},\ \bibinfo
  {pages} {1273} (\bibinfo {year} {2020})}\BibitemShut {NoStop}%
\bibitem [{\citenamefont {Olmon}\ \emph {et~al.}(2012)\citenamefont {Olmon},
  \citenamefont {Slovick}, \citenamefont {Johnson}, \citenamefont {Shelton},
  \citenamefont {Oh}, \citenamefont {Boreman},\ and\ \citenamefont
  {Raschke}}]{Olmon2012}%
  \BibitemOpen
  \bibfield  {author} {\bibinfo {author} {\bibfnamefont {R.~L.}\ \bibnamefont
  {Olmon}}, \bibinfo {author} {\bibfnamefont {B.}~\bibnamefont {Slovick}},
  \bibinfo {author} {\bibfnamefont {T.~W.}\ \bibnamefont {Johnson}}, \bibinfo
  {author} {\bibfnamefont {D.}~\bibnamefont {Shelton}}, \bibinfo {author}
  {\bibfnamefont {S.-H.}\ \bibnamefont {Oh}}, \bibinfo {author} {\bibfnamefont
  {G.~D.}\ \bibnamefont {Boreman}},\ and\ \bibinfo {author} {\bibfnamefont
  {M.~B.}\ \bibnamefont {Raschke}},\ }\href
  {https://doi.org/10.1103/PhysRevB.86.235147} {\bibfield  {journal} {\bibinfo
  {journal} {Phys. Rev. B}\ }\textbf {\bibinfo {volume} {86}},\ \bibinfo
  {pages} {235147} (\bibinfo {year} {2012})}\BibitemShut {NoStop}%
\bibitem [{COM()}]{COMSOL}%
  \BibitemOpen
  \href@noop {} {}\bibinfo {howpublished} {\url{www.comsol.com}}\BibitemShut
  {NoStop}%
\bibitem [{\citenamefont {Jackson}(1999)}]{JohnDavid1998}%
  \BibitemOpen
  \bibfield  {author} {\bibinfo {author} {\bibfnamefont {J.~D.}\ \bibnamefont
  {Jackson}},\ }\href@noop {} {\emph {\bibinfo {title} {Classical
  Electrodynamics, Third Edition}}}\ (\bibinfo  {publisher} {Wiley},\ \bibinfo
  {address} {New York},\ \bibinfo {year} {1999})\BibitemShut {NoStop}%
\bibitem [{\citenamefont {Fan}\ \emph {et~al.}(2003)\citenamefont {Fan},
  \citenamefont {Suh},\ and\ \citenamefont {Joannopoulos}}]{Fan2003}%
  \BibitemOpen
  \bibfield  {author} {\bibinfo {author} {\bibfnamefont {S.}~\bibnamefont
  {Fan}}, \bibinfo {author} {\bibfnamefont {W.}~\bibnamefont {Suh}},\ and\
  \bibinfo {author} {\bibfnamefont {J.~D.}\ \bibnamefont {Joannopoulos}},\
  }\href {https://doi.org/10.1364/JOSAA.20.000569} {\bibfield  {journal}
  {\bibinfo  {journal} {J. Opt. Soc. Am. A}\ }\textbf {\bibinfo {volume}
  {20}},\ \bibinfo {pages} {569} (\bibinfo {year} {2003})}\BibitemShut
  {NoStop}%
\bibitem [{\citenamefont {{Wonjoo Suh}}\ \emph {et~al.}(2004)\citenamefont
  {{Wonjoo Suh}}, \citenamefont {{Zheng Wang}},\ and\ \citenamefont {{Shanhui
  Fan}}}]{Suh2004}%
  \BibitemOpen
  \bibfield  {author} {\bibinfo {author} {\bibnamefont {{Wonjoo Suh}}},
  \bibinfo {author} {\bibnamefont {{Zheng Wang}}},\ and\ \bibinfo {author}
  {\bibnamefont {{Shanhui Fan}}},\ }\href
  {https://doi.org/10.1109/JQE.2004.834773} {\bibfield  {journal} {\bibinfo
  {journal} {IEEE J. Quantum Electron.}\ }\textbf {\bibinfo {volume} {40}},\
  \bibinfo {pages} {1511} (\bibinfo {year} {2004})}\BibitemShut {NoStop}%
\end{thebibliography}%
\end{document}